\documentclass{aa}  

\usepackage{graphicx}
\usepackage{txfonts}
\usepackage{xcolor}
\usepackage[para,online,flushleft]{threeparttable}
\usepackage{booktabs}
\usepackage{multirow}
\usepackage{enumitem}
\usepackage{hyperref}
\usepackage[normalem]{ulem}
\usepackage{mathrsfs}

\hypersetup{
  colorlinks= true,   
  urlcolor  = blue,     
  linkcolor = red, 
  citecolor = blue
}

\makeatletter
\renewcommand*\aa@pageof{, page \thepage{} of \pageref*{LastPage}}
\makeatother

\begin{document}

   \title{J-PAS: First Identification, Physical Properties and Ionization Efficiency of Extreme Emission Line Galaxies}

\author{
A.~Giménez-Alcázar\inst{1} \and 
R.~Amorín\inst{1} \and
J.~M.~Vílchez\inst{1} \and
A.~Hernán-Caballero\inst{2,3} \and
M.~González-Otero\inst{1} \and
A.~Arroyo-Polonio\inst{1} \and
J.~Iglesias-Páramo\inst{1} \and
A.~Lumbreras-Calle\inst{2,3} \and
J.~A.~Fernández-Ontiveros\inst{2,3} \and
C.~López-Sanjuan\inst{2,3} \and
L.~Bonatto\inst{4,5} \and
R.~M.~González~Delgado\inst{1} \and
C.~Kehrig\inst{1,6} \and
A.~Torralba\inst{7} \and
P.~T.~Rahna\inst{2} \and
Y.~Jiménez-Teja\inst{1} \and
I.~Márquez\inst{1} \and
I.~Breda\inst{8} \and
A.~Álvarez-Candal\inst{1} \and
R.~Abramo\inst{9} \and
J.~Alcaniz\inst{6} \and
N.~Benitez\inst{1} \and
S.~Bonoli\inst{10} \and
S.~Carneiro\inst{6} \and
J.~Cenarro\inst{2,3} \and
D.~Cristóbal-Hornillos\inst{2} \and
R.~Dupke\inst{6} \and
A.~Ederoclite\inst{2,3} \and
C.~Hernández-Monteagudo\inst{11} \and
A.~Marín-Franch\inst{2,3} \and
C.~Mendes~de~Oliveira\inst{12} \and
M.~Moles\inst{2} \and
L.~Sodré Jr.\inst{12} \and
K.~Taylor\inst{13} \and
J.~Varela\inst{2} \and
H.~Vázquez~Ramió\inst{2,3}
}

\institute{
Instituto de Astrofísica de Andalucía (IAA-CSIC), Glorieta de la Astronomía s/n, 18008 Granada, Spain\\ \email{agimenez@iaa.es}
\and
Centro de Estudios de Física del Cosmos de Aragón (CEFCA), Plaza San Juan 1, 44001 Teruel, Spain
\and
Unidad Asociada CEFCA-IAA, CEFCA, Unidad Asociada al CSIC por el IAA y el IFCA, Plaza San Juan 1, 44001 Teruel, Spain
\and
Vicerrectoría de Investigación y Postgrado, Universidad de La Serena, 1700000, Chile
\and
Istituto Nazionale di Geofisica e Vulcanologia, Via di Vigna Murata 605, 00143 Rome, Italy
\and
Observatório Nacional, Rua General José Cristino 77, São Cristóvão, 20921-400 Rio de Janeiro, Brazil
\and
Institute of Science and Technology Austria (ISTA), Am Campus 1, 3400 Klosterneuburg, Austria
\and
Dep. of Astrophysics, University of Vienna, Türkenschanzstraße 17, 1180 Vienna, Austria
\and
Instituto de Física, Universidade de São Paulo (IF/USP), São Paulo, Brazil
\and
Donostia International Physics Center (DIPC), Manuel Lardizabal Ibilbidea 4, San Sebastián, Spain
\and
Instituto de Astrofísica de Canarias (IAC), C/ Vía Láctea s/n, 38205 La Laguna, Tenerife, Spain; Universidad de La Laguna (ULL), Avda Francisco Sánchez, 38206 San Cristóbal de La Laguna, Tenerife, Spain
\and
Instituto de Astronomia, Geofísica e Ciências Atmosféricas, Universidade de São Paulo (IAG/USP), São Paulo, Brazil
\and
Instruments, 4121 Pembury Place, La Canada Flintridge, CA 91011, USA
}

   \date{Received -; -}

  \abstract
   {Extreme emission line galaxies (EELGs) are expected to have an important contribution to the star formation activity and mass assembly in galaxies. Specifically. EELGs are also promising candidates to have a leading role in the cosmic reionization as their interstellar medium may allow a significant fraction of their ionizing photons to escape ( > 5\%). Finding low-redshift analogues of these high-z galaxies is therefore essential to characterize the physical conditions in the ISM of these galaxies and understand the processes that reionized the Universe.}
   {We aim to develop a robust and efficient method for the photometric identification of EELGs using the J-PAS survey. J-PAS will cover approximately 8500 deg$^2$ of the sky with 54 narrow-band filters in the optical range plus $i$-SDSS, enabling detailed studies of the physical properties of these galaxies. In this work, we focus on an initial subset of the survey: a 30 square degree area with complete observations in all bands.}
   {We combine equivalent width (EW) measurements from J-PAS narrow-band photometry with artificial intelligence techniques to identify galaxies with emission lines exceeding 300 \AA\, in any emission line. We validate our selection using spectroscopic data from DESI/DR1 and characterize the selected sample through spectral energy distribution (SED) fitting with \texttt{CIGALE}.}
   {We identify 917 EELGs up to $z = 0.8$ over 30 deg$^2$, achieving a purity of 95\% and a completeness of 96\% for $i$-SDSS < 22.5 mag. Importantly, AGN contamination has been carefully considered and is estimated to be around 5\%. Furthermore, a cross-match with DESI yields 79 counterparts, whose redshifts are in excellent agreement with our photometric estimates, thereby confirming the reliability of our redshift determination. In addition, the derived emission line fluxes are in good agreement with spectroscopic measurements, reinforcing the robustness of our methodology. Moreover, the selected sample reveals strong correlations between ionizing photon production efficiency ($\xi_{\text{ion}}$) and EW(H$\beta$), which are consistent with previous observational studies at low and high redshift and theoretical expectations. Finally, most of the sources surpass the ionizing efficiency threshold required for reionization, highlighting their relevance as local analogues of early-universe galaxies. }
   {}


   \maketitle
%

\section{Introduction}
Over the past decades, several works have identified a population of galaxies undergoing powerful star formation events, known as Extreme Emission Line Galaxies (EELGs). These galaxies are characterized by an optical spectrum dominated by intense emission lines, such as  [\ion{O}{III}] $\lambda\lambda$4959, 5007 and H$\alpha$, with exceptionally high restframe equivalent widths (EWs), often reaching hundreds of angstroms (e.g., \citealp{vanderWel2011,Maseda2014ApJ...791...17M,Amorin2014A&A...568L...8A,Amorin2015A&A...578A.105A,calabro2017A&A...601A..95C,delmoral2024A&A...688A..28D}), and a blue continuum, generally bright in the UV but comparatively faint at optical wavelengths. EELGs are typically compact systems with stellar masses below 10$^9$ M$_\odot$ and exhibit elevated specific star formation rates (sSFRs) ranging from 10 to 100 Gyr$^{-1}$ \citep{Amorin2014A&A...568L...8A,calabro2017A&A...601A..95C,Tang2019MNRAS.489.2572T,arroyo2023A&A...677A.114A, Boyett2024MNRAS.535.1796B}.\\

Regarding their chemical abundances, EELGs are metal-poor systems, typically with subsolar metallicities ($\sim$20\% solar on average; e.g.,  \citealp{amorin2010ApJ...715L.128A,Amorin2014A&A...568L...8A,perezMontero2021MNRAS.504.1237P}). In the low-end of the metallicity distribution of EELGs, systems with the lowest oxygen abundances as measured from the direct Te-method, below a few percent solar, are found (e.g., \citealp{papaderos2008A&A...491..113P,morales2011ApJ...743...77M,kojima2020ApJ...898..142K}). Such extremely metal-poor EELGs are now routinely found at high redshifts with JWST (e.g., \citealp{cameron2023A&A...677A.115C, Llerena2024A&A...691A..59L,nakajima2024arXiv241204541N, laseter2024A&A...681A..70L,cullen2025MNRAS.540.2176C}). As such, they represent unique laboratories for studying star formation and ionized gas physics in nearly pristine environments. The number density of EELGs increases markedly with redshift (e.g., \citealp{smit2014ApJ...784...58S, Maseda2018ApJ...854...29M}), with up to an order of magnitude rise in the fraction of star-forming galaxies exhibiting extreme emission lines between $z \sim 2$ and $z \sim 7$ (\citealp{Boyett2022ApJ...940L..52B,Boyett2024MNRAS.535.1796B}). This is accompanied by a systematic increase in typical H$\alpha$ equivalent widths with redshift \citep{Stefanon2022ApJ...935...94S}. \cite{atek2014ApJ...789...96A} found that galaxies with H$\alpha$ equivalent widths greater than 300\,\AA\, contribute up to $\sim13\%$ of the total star formation rate density at $z \approx $ 1–2. This trend suggests that while EELGs are rare in the local Universe, they become increasingly common towards higher redshifts, and frequent in the reionization era. These galaxies produce notable amounts of photoionizing radiation, contributing significantly to the ultraviolet photon budget required for the reionization of the Universe (e.g., \citealp{Endsley10.1093/mnras/stad1919,naidu2022MNRAS.510.4582N,finkelstein2019ApJ...879...36F}).

Finding EELGs at low redshift is harder. \cite{perezMontero2021MNRAS.504.1237P} identified only about 2000 EELG candidates across the entire SDSS DR8, which corresponds to just 0.2\% of the total galaxy sample, highlighting their rarity in local surveys. This scarcity has been further confirmed in ongoing work using DESI (Amorín et al., in prep.), reinforcing the idea that EELGs represent a transient and uncommon phase at low $z$. Several factors may explain this rarity: the intrinsically low luminosity and low number density of EELGs, and observational selection effects (e.g., Malmquist bias, since EELGs tend to be low-mass systems and at higher redshifts only the most extreme ones can be detected at fixed stellar mass), as well as the short-lived nature of the EELG phase, which likely corresponds to brief, intense starburst episodes. Therefore, the detection and characterization of local EELG analogues open a unique window into understanding the physical processes that governed the formation and ionization conditions of galaxies during the reionization epoch.\\

EELGs in the local Universe exhibit a tremendous diversity and can fall into different categories based on their appearance and/or colour in images, the selection method used, and their redshift. Examples include, HII galaxies \citep[e.g.,][]{terlevich1991A&AS...91..285T,carol004AJ....128.1141K}, Blue Compact Dwarfs (BCDs; \citealp[e.g.,][]{Thuan1981ApJ...247..823T, Papaderos1996A&AS..120..207P}), Green Pea galaxies \citep[e.g.,][]{cardamone2009MNRAS.399.1191C, amorin2010ApJ...715L.128A, izotov2011ApJ...728..161I}, Blueberry Galaxies \citep[e.g.,][]{yang2017ApJ...847...38Y}, ELdots \citep[e.g.,][]{beki2015MNRAS.454L..41B}, as well as the recently discovered Little Red Dots \citep[e.g.,][]{labee2023Natur.616..266L, Kocevski2023ApJ...954L...4K, Matthee2024ApJ...963..129M, perez-gonzalez2024ApJ...968....4P}.
There are two well-established strategies for identifying these types of galaxies. The first involves spectroscopic surveys, where EELGs are typically recognized by their intense emission lines with unusually large EWs, resulting from the photoionization of the gas by hot massive stars. For example, \cite{Amorin2015A&A...578A.105A} reported 165 EELGs in a 1.7 square degree in zCOSMOS 20k-Bright spectroscopic survey. 

The second procedure relies on photometric surveys, which detect flux excess in a specific band compared to adjacent bands or through colour excess. Narrow-band photometry has proven effective in identifying EELGs, as shown by \cite{iglesias2022A&A...665A..95I}, who found 17 EELGs in the miniJPAS area, and \cite{lumbreras2022A&A...668A..60L}, who reported 466 in the J-PLUS survey. Medium- and broad-band photometry have also been widely used; for instance, \cite{vanderWel2011} identified 70 EELGs in the CANDELS fields, and \cite{Withers2023ApJ...958L..14W} selected 118 EELGs at $1.7 < z < 6.7$ using JWST color criteria. Continuing with JWST data, \cite{Llerena2024A&A...691A..59L} analyzed $\sim$730 EELGs at $4 \lesssim z < 9$ from the CEERS survey (see also \citealp{Davis2024ApJ...974...42D,Boyett2022ApJ...940L..52B}). Other large samples have been identified through color-excess selection and spectroscopic follow-up, such as in \cite{Onodera2020ApJ...904..180O} at $z \sim 3.3$ and \cite{Tang2019MNRAS.489.2572T} at $1.3 < z < 2.4$, where many EELGs exhibit high O32 ratios and ionizing photon production efficiencies consistent with Lyman continuum leakage \citep{Jaskot2013ApJ...766...91J,Nakajima2014MNRAS.442..900N}.

A complete and consistent census of EELGs across a wide redshift range remains an observational challenge. To overcome it, we make use of the Javalambre-Physics of the Accelerating Universe Astrophysical Survey (J-PAS; \citealp{Benitez2014arXiv1403.5237B, Bonoli2022eas..conf.2468B}, Vázquez Ramiró et al. in prep) . J-PAS will observe a giant portion of the sky (8500 deg$^2$), with no target selection bias, helped by 54 narrow band filters in the optical regime with an average full width at half maximum (FWHM) of 145 \AA\, , spanning from 3780 \AA\ to 9100 \AA\,  and an average spatial resolution of FWHM < 1.5”. We aim to carry out a large-scale search for EELGs within J-PAS, as the first data releases are already available. The limit in EW for considering a galaxy as an EELG is not clearly defined and in general, the rest-frame EW evolves with redshift as ionization efficiency increases \citep[e.g.,][]{Sobral2014MNRAS.437.3516S,Khostovan10.1093/mnras/stw2174}. In this work we adopt the same criteria given in \cite{iglesias2022A&A...665A..95I} and \cite{breda2024MNRAS.528.3340B}. We consider EELGs as galaxies with rest frame $\mathrm{EW} \geq 300$\,\AA\, in at least one of the emission lines: \ion{O}{III} or H$\alpha$, and we adopt this criterion for the search. This criterion ensures that, according to stellar population models (e.g., Starburst99; \citealp{Leitherer1999ApJS..123....3L} and \citealp[]{pystarburst2025ApJS..280....5H}), the galaxy's spectrum is dominated by the light from a starburst younger than 10 Myr.\\

The structure of this paper is as follows. Sect. \ref{section2} presents the observational data. In Sect. \ref{section3}, we detail the methodology used to select the EELGs. Sect. \ref{section4} provides a characterization of the EELG sample. In Sect. \ref{section5}, we discuss the implications of EELGs in cosmic reionization. Conclusions are given in Sect. \ref{section6}.
In the following, we adopt the $\Lambda$CDM cosmological model with parameters $H_0 = 70\,\mathrm{km\,s^{-1}\,Mpc^{-1}}$, $\Omega_{M} = 0.3$, and $\Omega_{\Lambda} = 0.7$.

\section{Data}
\label{section2}
\begin{figure}
    \centering
    \includegraphics[width=1\linewidth]{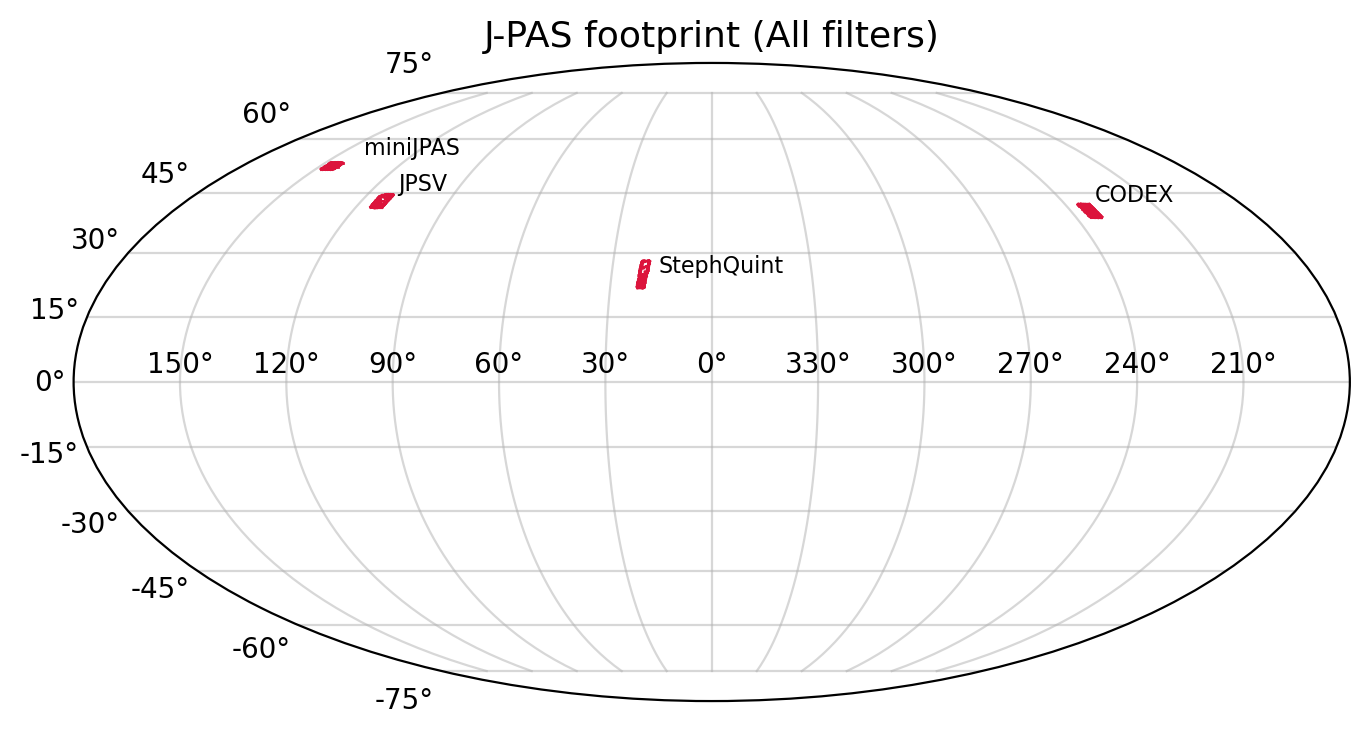}
    \caption{J-PAS IDR202406 observed footprint with all the filters showing the positions of the seed fields. 
The coordinates (RA, DEC) in degrees are: 
CODEX (126.1125, 40.1053), 
miniJPAS (214.4500, 52.7261), 
JPSV (244.00, 43.00), 
and StephQuint (339.00, 22.50).}
    \label{path}
\end{figure}

This work makes use of the first internal Data Release of J-PAS (IDR202406) covering a total field of view of approximately 380 deg$^2$ conducted at the Observatorio Astrofísico de Javalambre (OAJ) at Teruel, Spain . It is composed of 54 narrow-band filters covering the 3780–9100 \AA\, wavelength range, with a full width at half maximum (FWHM) of $\approx$ 145 \AA\,, equally spaced every 100 \AA\,. In addition, two intermediate-band filters cover the UV edge (uJAVA, $\lambda_c$ = 3497 \AA\,, FWHM = 495 \AA\,) and the red edge (J1007, $\lambda_c$ = 9316 \AA\,, FWHM = 620 \AA\,). The achieved spectral resolution is R $\approx$ 60, equivalent to very low-resolution spectroscopy. A detailed description of the observations, telescope, and instrumental setup can be found in \cite{Mar2024SPIE13096E..1QM}. For our study, we include only sources with available measurements in all bands, excluding any incomplete observations. This selection reduces the effective survey area to 29 square degrees. Source identification in the photometric images was carried out using SExtractor \citep{Bertin1996A&AS..117..393B} with two different approaches: single-mode and dual-mode. In dual-mode, object detection and aperture definition are performed using the $i$-SDSS band as a reference with a magnitude limit of 23.5 mag in AB system. Photometry in all filters is then measured within the same predefined apertures ($i$-SDSS aperture) and exactly in the same coordinates, ensuring consistent flux extraction across bands. On the other hand, single-mode detection treats each band independently, identifying sources separately in each image without imposing a reference aperture.  Here, we strictly use the catalogue produced in dual mode, what ensures that all sources have a common processing frame based on $i$-SDSS band, enabling consistent measurements across all bands. Figure~\ref{path} shows the four sky patches observed in the J-PAS IDR, along with the region covered by miniJPAS \cite{bonoli}, which was used in previous studies (e.g., \citealp{gonzalez2021A&A...649A..79G,gines2022A&A...661A..99M,torralba2023A&A...680A..14T}). However, photometric redshifts are only computed for sources with AB magnitude brighter than 22.5 in the $i$-SDSS band. It is worth noting that the narrow-band observations are generally shallower than the $i$-band data, with 5$\sigma$ limiting magnitudes ranging between 21.0 and 22.4, and a median depth of 21.9 (Hernán-Caballero et al., in prep). Therefore, to ensure a reliable characterization of the resulting population, our final sample will be limited to sources with $i$-SDSS < 22.5.\\

The observations were reduced using a dedicated pipeline developed at Centro de Estudios de Física del Cosmos de Aragón (CEFCA) for processing images from the two main telescopes of the OAJ, JST250 and JAST80. This data reduction pipeline has already been successfully used in the processing of previously released datasets, such as J-PLUS (\citealp{cenarro,lopez-sanjuan}) and miniJPAS \citep{bonoli}. Photometric calibration (Vázquez-Ramiró et al. in .) was done using the Gaia data release 3 \citep{gaia2023A&A...674A...1G}.

\section{Methodology}
\label{section3}
Identifying EELGs presents a challenge due to their low number density and the growing volume of data, which make traditional methods inefficient. With J-PAS, we can now adopt a more powerful approach, taking advantage of both the spectrophotometric information and the 54 narrow-band images available for each source. In this work, we propose combining a traditional search method with a machine learning approach to efficiently handle large datasets and improve the identification of EELGs in current and future data releases.

\subsection{Step I: Classical search}

The J-PAS catalogue contains a vast number of objects that need to be filtered. To begin the process, we have designed an operational flow that ensures specific criteria are met to determine whether or not a source should be included. The photospectrum is first processed by removing emission lines or artifacts using 3-sigma clipping, followed by a polynomial fit (of 1st and 2nd degrees, with the best fit chosen based on root mean square error) to model the continuum. This step is repeated three times, taking into account the uncertainties in each data point: once for the upper limit, once for the lower limit, and once for the standard data points. In this way, we ensure a robust measurement of the continuum level for galaxies where it is reliably detected. \\

To avoid false detections driven by noise, we require that an emission line must have a flux at least 5$\sigma_{\text{cont}}$ above the continuum, where $\sigma_{\text{cont}}$ represents the standard deviation of the continuum level in the vicinity of the line. This criterion ensures that lines are not buried in the spectral noise and helps avoid spurious detections in noisy spectra. While this threshold increases the reliability of the line detection and the measurement EWs, it may limit our ability to identify EELGs in cases where the continuum is too faint or not detected with sufficient confidence. However, this trade-off is necessary to ensure that derived EWs are not artificially boosted by an underestimated or noisy continuum. Next, following \cite{iglesias2022A&A...665A..95I}, we apply the contrast factor to select only those sources EW greater than 300 \AA. The criterion is defined as

\begin{equation}
    F_L(\lambda) - F_C(\lambda) \geq \frac{EW}{EW + \Delta_F} \times F_L(\lambda),
\end{equation}where EW is 300 \AA \ and $\Delta_F$ is the filter width. The choice for 300 \AA\, facilitates the identification of young, low-metallicity galaxies undergoing intense starburst episodes. Other minor conditions are: a signal-to-noise ratio lower than 8 in individual measurements, and fluxes in the band lower than $3 \times 10^{-15}~\mathrm{erg\,s^{-1}\,cm^{-2}\,\text{\AA}^{-1}}$ or negative are discarded. We do not apply the typical star/galaxy separation because EELGs are often extremely compact, and classifiers can sometimes become confused with stars (e.g. \citealp{cardamone2009MNRAS.399.1191C}), leading to misclassification.

\subsection{Step II: AGN/Quasar and cosmetic defects}

After applying Step I, we expect to obtain a clean sample of galaxies with well-defined emission lines and EW greater than 300\,\AA. However, some contaminants may still remain in the sample. For instance, there are cases where the polynomial continuum fit (of first or second degree) fails, leading to an incorrect continuum estimation, such as in star-like spectra. Additionally, certain objects may show spurious emission features due to contamination in a specific band, for example by diffraction spikes which can artificially boost the flux and mimic an emission line. Moreover, quasi stellar objects (QSOs) can also appear in the sample due to their strong emission lines; however, our goal is to identify emission lines produced by star formation, not those associated with QSO activity. To minimize these sources of contamination, we trained a neural network to distinguish EELGs from other types of objects.\\

\subsubsection{Neural Network: input and output}
We normalize the photospectra to a range of [0, 1] by dividing each spectrum by its maximum flux value. This normalization is done independently for each object, and thus the photospectra are normalized with respect to their own maximum. This procedure ensures consistent input scaling for the neural network (NN), but it does not preserve absolute flux information. As such, the normalization is relative (not physically calibrated) and is primarily intended to facilitate efficient NN training. The training sample consists of 684 sources labelled as EELGs, carefully selected to build a representative and reliable EELG dataset. These galaxies were initially identified using the photometric selection method described in \cite{iglesias2022A&A...665A..95I}. Each candidate was then visually inspected to confirm its nature and ensure it met the established EELG selection criteria mentioned above. It is precisely during this manual inspection step that the need for automated methods becomes evident: while feasible for a few hundred sources, visual confirmation quickly becomes impractical when scaling up to the full J-PAS dataset. In addition to labelled EELGs, the training set includes 1160 sources labelled as non-EELGs. This group comprises a diverse set of objects, including spectroscopically confirmed QSOs, obtained by cross-matching J-PAS with the DESI/DR1 catalogue \citep{desi2025arXiv250314745D}, as well as a variety of normal galaxies, such as elliptical, late-type, and spiral galaxies. These non-EELG sources serve as a contrasting population that enables the model to learn to distinguish EELGs from other types of galaxies and active objects. As the classes in the training set are not balanced, we apply class weights inversely proportional to the number of instances in each class to mitigate bias during training. Additionally, 70\% of the dataset is used for training, while the remaining 30\% is reserved for testing and performance evaluation, remaining unseen by the network during training.\\

To quantify the likelihood of a galaxy being an EELG, our primary neural network outputs an ``EELGness'' score, denoted as $P_0$, ranging from 0 to 1, where higher values indicate a stronger probability of being a true EELG. This network processes both the photometric spectrum and the galaxy image, following the architecture shown in Figure~\ref{fig:model}, but considering only the upper and middle branches (spectral and image processing). However, during visual inspection of the galaxy images, we identified a significant number of false positives caused by cosmetic defects artifacts in the imaging data that produce emission-line-like features but do not correspond to real astrophysical objects. To address this, we developed a second neural network, specifically trained to distinguish between true EELGs and these cosmetic artifacts. This network outputs a complementary score, $P_1$, which reflects the probability that a given detection is a genuine EELG and not a cosmetic defect. Both networks share the same fundamental architecture, with one key difference: the cosmetic-defect classifier includes an additional dedicated branch (Figure~\ref{fig:model}, lower branch) designed to analyse image regions associated with the location of the emission line, improving the network's ability to recognize defects.
For the standard EELG classifier ($P_0$), only the first two branches (spectral and image) are active. For the cosmetic-defect classifier ($P_1$), all three branches are used, including the defect-detection branch. In both cases, the outputs of the active branches are concatenated and passed through a common set of dense layers, which integrate the extracted features and produce the final classification output. This design ensures that both spectral and morphological information, as well as potential image artifacts, are jointly considered for more reliable EELG identification. A deep description of the architecture and training of the NN is shown in appendix \ref{ApendixNN}.

\begin{figure*}
    \centering
    \includegraphics[width=0.9\linewidth]{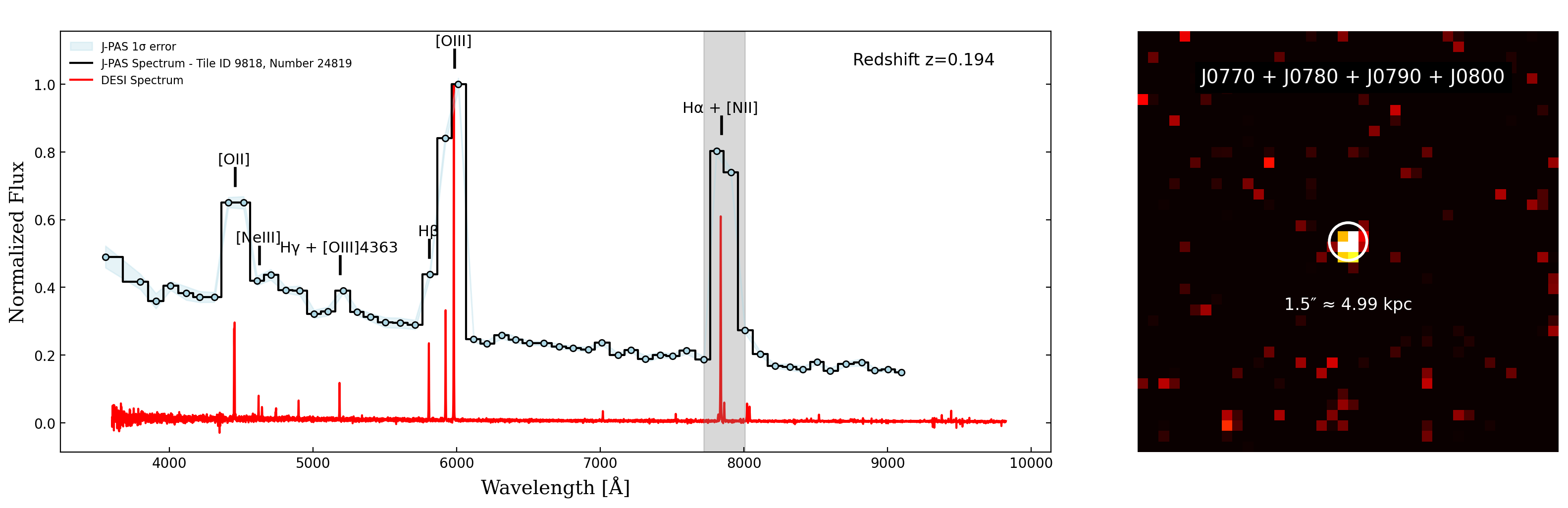}
\end{figure*}

\begin{figure*}
    \centering
    \includegraphics[width=0.9\linewidth]{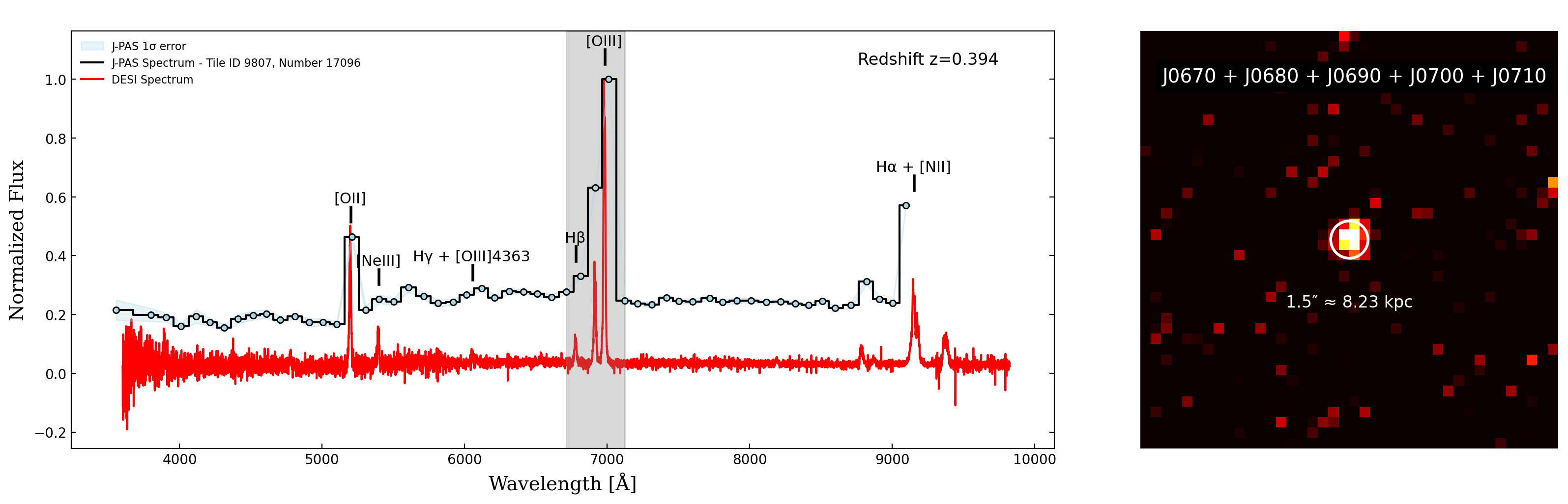}
    \label{fig:espctros}
\end{figure*}

\begin{figure*}
    \centering
    \includegraphics[width=0.9\linewidth]{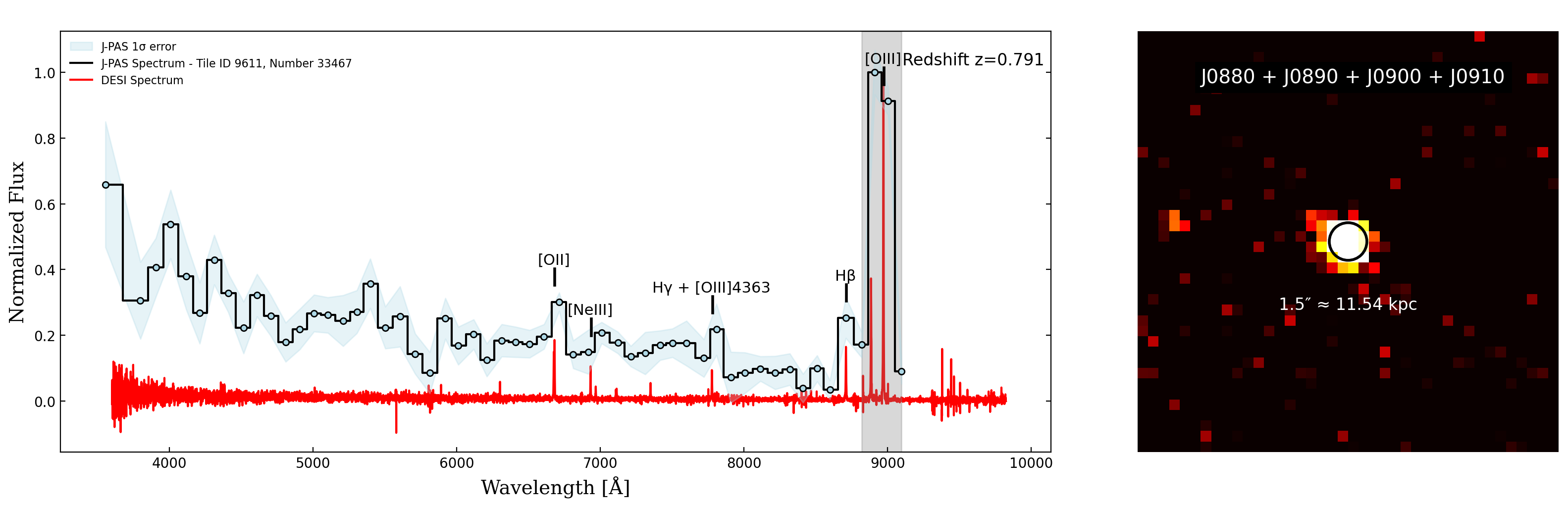}
    \caption{Data products from J-PAS for the EELG candidates. Left: The black line represents the J-PAS photometric spectrum, while the red line shows the corresponding DESI spectrum. The shaded gray region marks the wavelength range selected for integration. Right: Image cutouts resulting from integrating the data cube over the selected spectral region. The horizontal white bar corresponds to 2 arcseconds in length. Highlight that with J-PAS we are able to detect the continuum, whereas DESI cannot.}
    \label{espectros}
\end{figure*}

\section{Sample selection and characterization}
\label{section4}
We selected galaxies with $P_0 \geq 0.8$, ensuring a well-balanced trade-off between completeness and purity. This threshold allows us to recover nearly 97\% of extreme emission line emitters in the test sample, as determined from the classifier’s performance on a labelled test set. Applying $P_1 > 0.2$ reduces the sample from 969 to 917 galaxies, eliminating possible fake detections in the images. Of the final sample, 79 galaxies have counterparts in the latest release of the DESI/DR1 catalogue (Fig.\ref{espectros}). Assuming that the DESI selection is not strongly biased for or against EELGs, we can assess the completeness of our selection by examining how many DESI sources with EW greater than 300 \AA\, in at least one of the emission lines, fall within the J-PAS footprint and among those, how many are successfully recovered by our method. A total of 28 DESI EELGs sources lie within the J-PAS footprint. Eight out of 28 are not present in the J-PAS catalogues: seven correspond to starburst regions in large spiral galaxies, and one object with an I-band magnitude fainter than 23.5 which exceeds the J-PAS limiting value. Among the 20 DESI EELGs within the J-PAS catalogue, 16 are successfully recovered by our selection methodology. The remaining four galaxies are not detected because they do not satisfy the EW criteria in J-PAS. This is the case of some cometary/tadpole BCDs (e.g. \citealp{papaderos2008A&A...491..113P,almeida2015ApJ...810L..15S}), in which the emission line EW of the compact starburst spectrum captured by the DESI fibre is different from the EW measured by the J-PAS aperture photometry, which in those cases includes part of the faint cometary/tadpole tail with a less extreme spectrum, i.e. fainter nebular contribution. Additionally, DESI targets compact sources by placing the spectroscopic aperture on the brightest knot within the galaxy, which often results in high EW measurements. In contrast, when the photometric measurement is integrated over the entire galaxy, the EW can be significantly diluted due to the contribution of the older, underlying stellar population. This distinction arises from the very definition of an EELG: in our approach, we do not define an EELG as a galaxy that merely hosts a burst, but rather as a galaxy in which the burst dominates the optical luminosity, outshining the additional contribution of older stellar populations from the host galaxy (e.g. \citealp{amorin2009A&A...501...75A, amorin2012ApJ...749..185A,fernandez2022MNRAS.511.2515F}). Following our selection criteria, we have recovered all the EELGs within the J-PAS footprint, with the exception of the faint object.

\subsection{Purity and AGN contribution}

\begin{figure}
    \centering
    \includegraphics[width=1\linewidth]{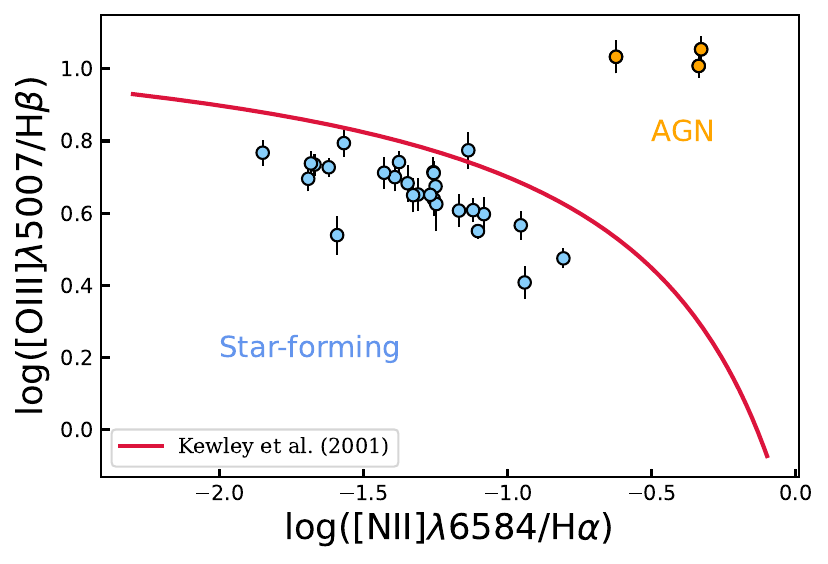}
    \caption{BPT diagram (\citealp{BPT1981PASP...93....5B}). Blue point represent the data points that have a counterpart in DESI with the H$\alpha$. The data points classified as AGNs are plotted in orange. The solid red line corresponds with the Kewley relationship \citep{kewley2001ApJ...556..121K}.}
    \label{fig:bpt}
\end{figure}

 To assess AGN contribution, we employ the BPT diagram with the sources that have a counterpart in DESI/DR1 (Fig.\ref{fig:bpt}). Emission lines of the DESI spectra were measured with the LIME tool \citep{lime2024A&A...688A..69F}, revealing three AGN candidates showing broad components in Balmer lines. Additionally, we identified two galaxies exhibiting [\ion{Ne}{V}] $\lambda$3426 emission, a reliable AGN tracer. Although rare, [\ion{Ne}{V}]  has also been detected in low-metallicity star-forming galaxies \citep{izotov2021MNRAS.508.2556I,Mingozzi2025ApJ...985..253M, arroyo2025ApJ...987L..36A}. This brings the total AGN contamination to 5 out of 79 galaxies, approximately 6$\%$. Theses sources were removed from the sample. As it will be shown later (Sec \ref{Sec4.3}, Fig.\ref{fig:compare_cigale}), both the spectroscopic and J-PAS samples exhibit similar properties, suggesting that the overall AGN contamination in our selection is around 6$\%$.

\subsection{Redshift estimation}

In J-PAS photometric redshifts are computed for all sources with $i < 22.5$ using galaxy templates regardless of their morphological classification. This approach means that the resulting redshift probability distributions, $P(z)$, are conditional on the object being a galaxy. Such a method allows for the calculation of galaxy number densities that incorporate uncertainties in morphological classification (\citealp{hernar2021A&A...654A.101H}), and in our case, enables the construction of a complete EELG sample unbiased by morphology. 

We use a customized version of LePhare \citep{lephare2011ascl.soft08009A}, keeping the same configuration as used in the official photo-$z$ runs for the miniJPAS PDR \citep{hernar2021A&A...654A.101H} and J-NEP \citep{hernan2023A&A...671A..71H}, except for the addition of two specific EELG templates, which are described below. LePhare employs filter transmission profiles to compute synthetic photometry as a function of redshift for a given set of templates. EELGs, being rare and peculiar objects, require dedicated templates with detailed nebular emission features to accurately determine their redshift. We constructed spectroscopic templates based on a selection of AGN and star-forming galaxies from the DESI EDR.\\
 
\begin{figure}
    \centering
    \includegraphics[width=1\linewidth]{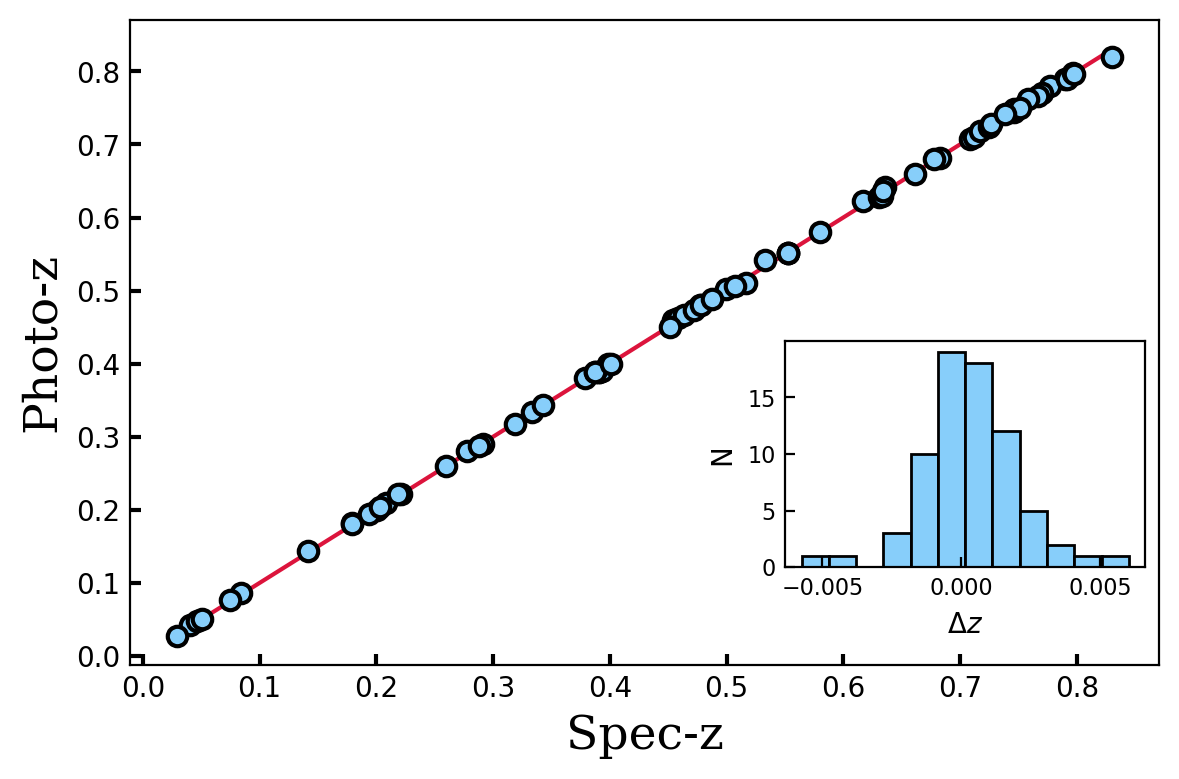}
    \caption{A 1:1 correlation is observed between the spectroscopic redshifts and the best-fit values from LePhare. The inset quantifies the relative differences between the two redshift estimates.
}
    \label{fig:z_spec}
\end{figure}

These templates span a wavelength range from 1400\,\AA\ to 10000\,\AA, covering the wavelength range of J-PAS filters set until redshift $\approx$ 0.9. They include representative SED of EELGs, particularly those exhibiting strong [O\,\textsc{iii}] and H$\alpha$ emission. This inclusion is crucial for accurate photometric redshift estimation, as standard template libraries used in photo-z codes like LePhare often lack examples of galaxies with such intense emission lines. To build these templates, we performed a stacking analysis of over 12997 star-forming galaxies and 350 AGN selected from the DESI spectroscopic sample. This approach allowed us to construct realistic SEDs that capture both the continuum and emission line features characteristic of EELGs. Excluding these templates slightly degrades the overall agreement between photometric and spectroscopic redshifts, particularly in cases where the algorithm confuses [O\,\textsc{iii}] with H$\alpha$. Further details on the DESI sample selection and template construction will be provided in Bonatto et al. (in prep.). Using these optimized templates, we find excellent agreement between the best-fit photometric redshifts from LePhare and the spectroscopic redshifts across the full redshift range up to z $\approx$ 0.9, as illustrated in Fig~\ref{fig:z_spec}. To quantify the photometric redshift accuracy, we compute the normalized median absolute deviation ($\sigma_{\mathrm{NMAD}}$), obtaining a value of 0.0015, five times lower than that obtained without using the templates.

Moreover, the redshift distribution of our sample is remarkably homogeneous (Appendix \ref{fig:z}), ensuring that our analysis is not biased by redshift-dependent effects. As an additional diagnostic, we examine the template type preferred by the redshift fitting. We find that 5\% of the galaxies are better fitted by an AGN template, which is fully consistent with our previous estimate of AGN contamination in the sample.

\subsection{SED fitting}

\label{Sec4.3}
In order to compute the physical properties, we make use of Code Investigating GALaxy Emission \citep[\texttt{CIGALE v2025.0};][]{cigale2019A&A...622A.103B}. \texttt{CIGALE} uses parametric star formation histories (SFHs) and outputs physical parameters and their uncertainties by using a Bayesian approach. It creates a probability distribution function for each parameter by evaluating the $\chi^2$ over the full set of models used for the fit. In this work, the SFH is modelled, as in previous works \citep[e.g.,][]{lumbreras2022A&A...668A..60L}, using a double exponential with an old population selected to represent the underlying galaxy, and a young one to reproduce the strong starburst, inducing the extreme emission lines. In addition to the J-PAS photometric points, we incorporated complementary photometric data covering the UV and IR regimes. Specifically, we included the FUV and NUV bands from the \textit{GALEX} telescope; the u,g,r,i,z bands from the Sloan Digital Sky Survey (SDSS); the \textit{WISE} bands at 3.4, 4.6, 12.1, and 22.2 $\mu$m; the \textit{Spitzer}/IRAC bands at 3.6, 4.5, 5.8, and 8 $\mu$m; and the \textit{Spitzer}/MIPS bands at 24 and 70 $\mu$m. For the stellar population synthesis, we have used the Charlot \& Bruzual (2019) models with the \cite{chabrier2003PASP..115..763C} initial mass function. Nebular emission is incorporated using the CLOUDY models \citep{Ferland2013RMxAA..49..137F}, and we assume a $f_\mathrm{esc}$ = 0.0, meaning that all Lyman continuum (LyC) photons are reprocessed into the Balmer lines, $f_\mathrm{dust}$ = 0 (no LyC absorbed by dust) and an electron density $n_e$ = 100 cm$^{-3}$. The attenuation in the models was taken into account by using the modified \cite{calcetti2000ApJ...533..682C} law and for the dust emission we used the models by \cite{dale2014ApJ...784...83D}. These extended photometric data improve the constraints on the SEDs, particularly in the UV, where they help characterize recent star formation, and in the IR, where they are essential to account for dust emission and reprocessed light. All the parameters employed in the fitting process are summarized in Table~\ref{tab:sed_parameters}.

\begin{figure}
    \centering
    \includegraphics[width=1\linewidth]{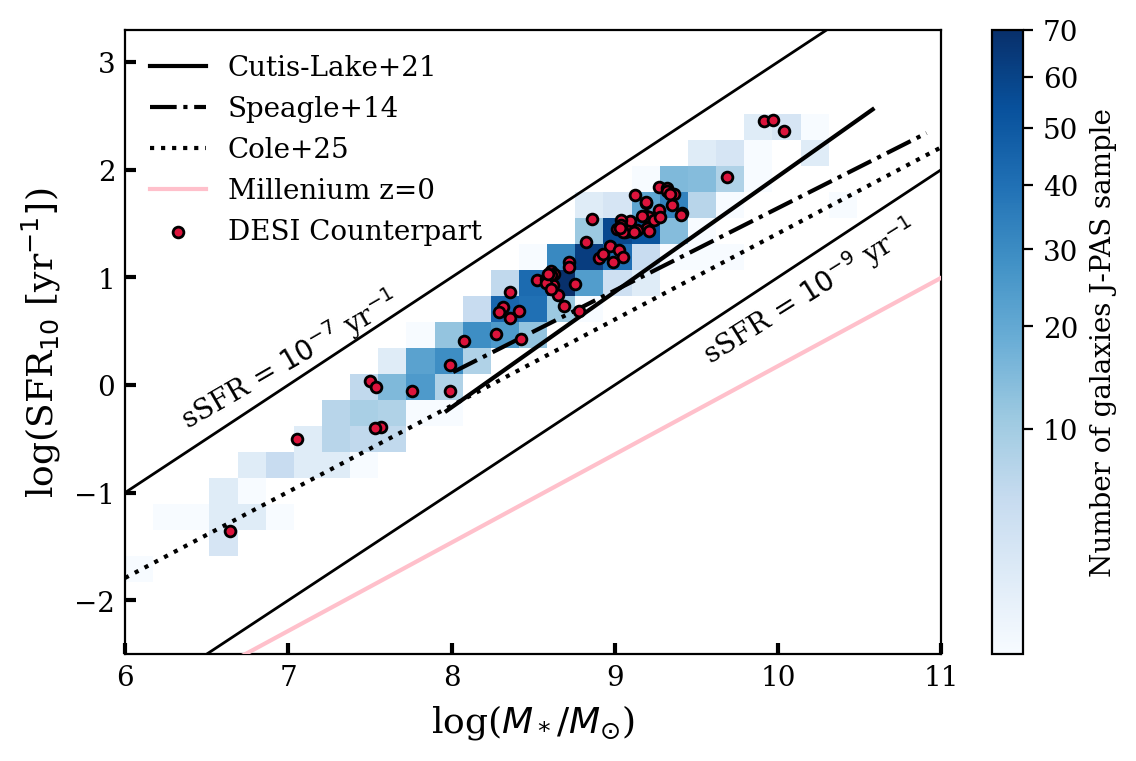}
    \caption{Main sequence of star-forming galaxies. The plot shows the logarithm of the star formation rate, $\log(\mathrm{SFR}_{10})$, versus the logarithm of the stellar mass, $\log(M_\star)$. The SFR$_{10}$ refers to the average star formation rate over the past 10 Myr. Red points indicate galaxies with spectroscopic observations from DESI. The dotted black line shows the relation from \citet{cole2025ApJ...979..193C} at redshift 4.5–5. The solid black line corresponds to the relation from (\citealp{curtis2021MNRAS.503.4855C}; mock photometric samples of galaxies at z $\approx$ 5), the dot-dashed line represents the relation from \citet{speagle2014ApJS..214...15S} (64 measurements of the star-forming "Main Sequence" from literature out to z $\approx$ 6) and the pink line shows the results from the Millennium Simulation \citep{millenium2005Natur.435..629S}. The dashed lines draw regions of constant specific star formation rate (sSFR) at values of -7 and -9.}
    \label{fig:MS}
\end{figure}

We estimate the continuum, stellar mass, star formation history and extinction, in order to properly characterize the sample. In Fig. \ref{fig:MS}, the main sequence (MS) of star forming galaxies is shown. Notably, our selected sample of EELGs is composed of galaxies with high star formation rates, above the star-forming main sequence at redshift $z \sim 4.5$--$5$,  derived using \textsc{BAGPIPES} (\citealp{bagpipes2018MNRAS.480.4379C}) in \cite{cole2025ApJ...979..193C}. This result assumes a \citet{chabrier2003PASP..115..763C} initial mass function  and applies the \citet{calcetti2000ApJ...533..682C} dust attenuation law. This was expected, as we selected galaxies with high EW, a characteristic feature of bursty galaxies, where their last burst of star formation occurs within the last 10 Myr.\\

\begin{figure}
    \centering
    \includegraphics[width=1\linewidth]{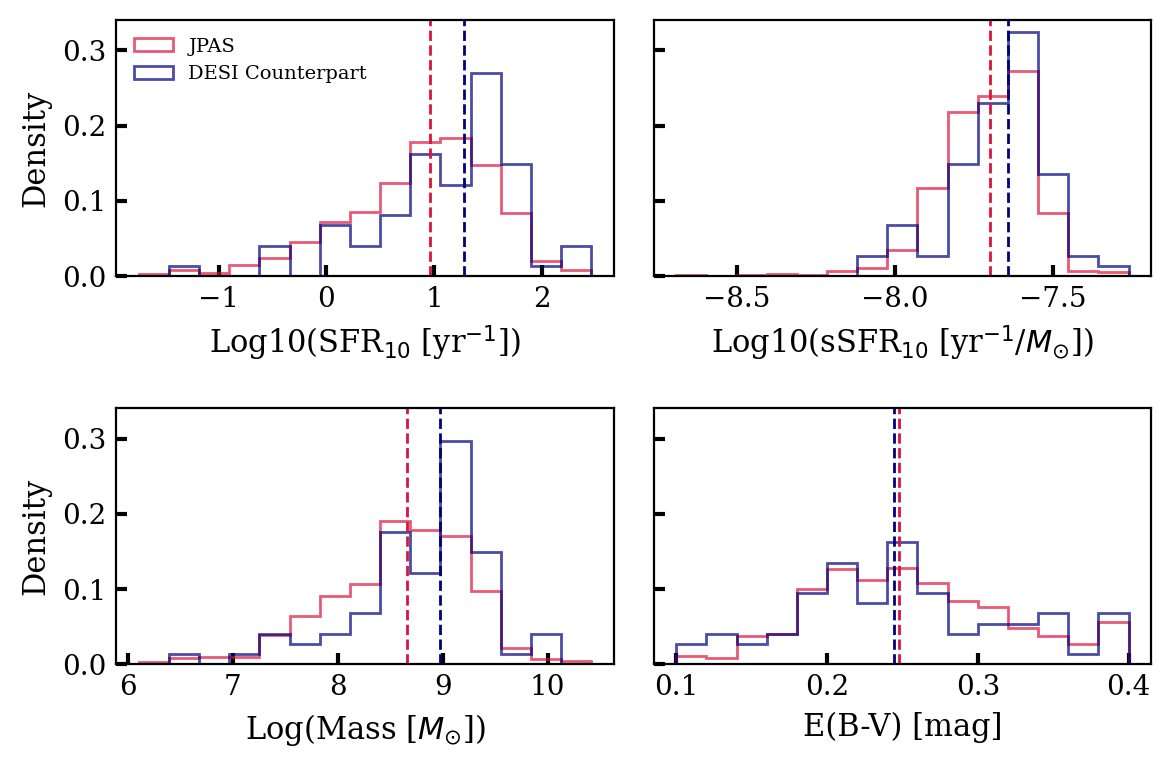}
    \caption{Histogram of the star formation rate (SFR$_{10}$, specific star formation rate (sSFR$_{10}$), stellar mass ($M_{\odot}$), and extinction (E(B-V)) for the J-PAS sample and its DESI counterparts. Vertical dashed lines indicate the mean values of each distribution.
}
    \label{fig:compare_cigale}
\end{figure}

Most galaxies fall within the $7<\log(M_\star/M_\odot)<10$ with a median of $\log(M_{\star}/M_{\odot}) = 8.66 \pm 0.02$. In comparison, the spectroscopic subsample yields a median mass of $\log(M_{\star}/M_{\odot}) = 8.98 \pm 0.09$. The spectroscopic counterpart appears slightly burstier (SFR$_{10}$), with a mean SFR of 0.97$\pm$ 0.02  M$_{\odot}$ yr$^{-1}$ in the full sample and 1.27 $\pm$ 0.13 M$_{\odot}$ yr$^{-1}$ in the Desi subsample. Moreover, most galaxies show relatively low dust extinction, with a median colour excess parameter of E(B-V)$_\mathrm{J-PAS}$= 0.25 $\pm$ 0.01 and E(B-V)$_\mathrm{DESI}$= 0.22 $\pm$ 0.01. To assess whether the two populations can be statistically considered equal, we performed a Kolmogorov-Smirnov test with 1000 Monte Carlo (MC) draws to account for uncertainties. The results indicate that although the SFR and stellar mass distributions show small but statistically significant differences (p < 0.05), the specific SFR over the recent 10 Myr (sSFR${10}$) and E(B-V), shows no significant deviation between JPAS and spectroscopy sample (p > 0.05). This indicates that the relative scaling between SFR and mass is preserved, and both samples are consistent with representing the same underlying galaxy population in terms of star-formation activity (see Fig.~\ref{fig:compare_cigale}). The mild differences that we can appreciate, might be attributed to the fibre selection function in DESI. 

\begin{table*}
    \tiny
    \centering
    \caption{Derived SED parameters for selected galaxies.}
    \begin{tabular}{cccccccccc}
        \toprule
        \toprule
        Tile ID & Number & RA & DEC & 
        Redshift & $R_{\rm eff}(\farcs)$ &
        $\log(\mathrm{M_{\star}}/\mathrm{M}_\odot)$ & $\log(\text{SFR}_{10}/M_{\odot} \text{yr}^{-1})$ &
        E(B$-$V) [mag]\\
        \midrule
        8838 & 7985  & 338.5187 & 22.7106 & 0.17  & 0.30 & 7.69 $\pm$ 0.37 & -0.15 $\pm$ 0.54 & 0.19 $\pm$ 0.09  \\
        8838 & 13110 & 338.9928 & 22.6274 & 0.46  & 0.27 & 9.51 $\pm$ 0.12 & 2.00 $\pm$ 0.76 & 0.30 $\pm$ 0.01  \\
        8838 & 15288 & 338.5909 & 22.5915 & 0.398 & 0.41 & 8.71 $\pm$ 0.39 & 0.91 $\pm$ 0.65 & 0.35 $\pm$ 0.07  \\
        8838 & 18374 & 339.1521 & 22.5362 & 0.586 & 0.35 & 9.11 $\pm$ 0.37 & 1.40 $\pm$ 0.48 & 0.32 $\pm$ 0.08  \\
    ...  & ...   & ... & ... & ... & ...             & ...             & ... & ...  \\
        \bottomrule
    \end{tabular}

    \vspace{2mm}
\begin{minipage}{\textwidth}
\small
\textbf{Notes.} 
The entire version of this table for the full sample of EELGs is available at the CDS.
\end{minipage}
\end{table*}

\subsection{Emission line flux extraction}

\begin{figure}
    \centering
    \includegraphics[width=.9\linewidth]{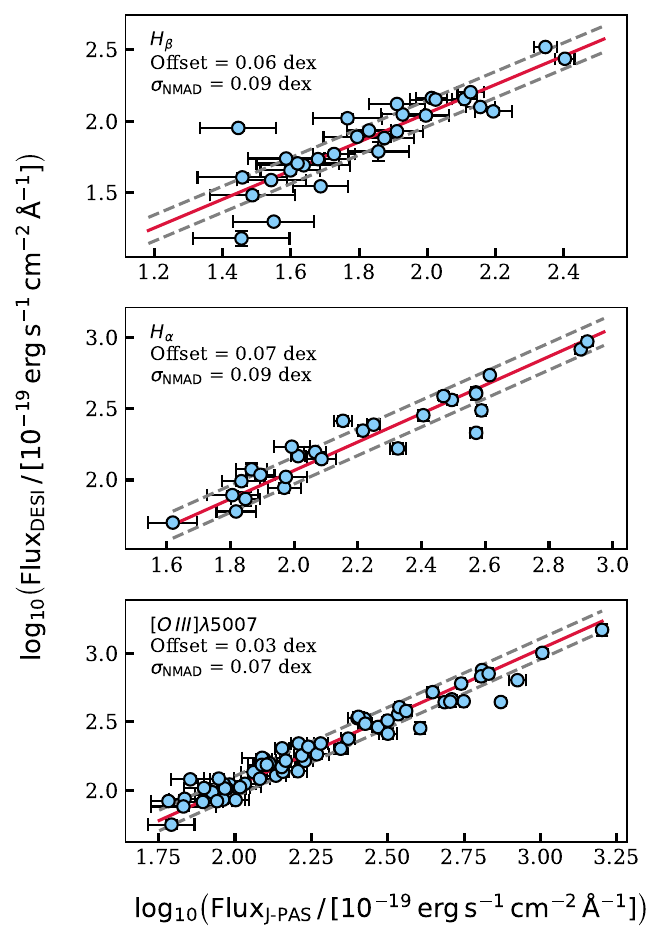}
    \caption{Photometric fluxes compared with the spectroscopic fluxes from the DESI counterpart for H$\beta$, H$\alpha$ and [\ion{O}{III}] 5007\,\AA\ from top to bottom. Grey lines indicates de limits of the $\pm1\sigma$ region. Red line represents the 1:1 ratio.}
    \label{fig:flux}
\end{figure}

One of the main spectral features of EELGs is their strong emission lines, thus capturing precise and robust measurements of line intensities is essential. Measuring fluxes using photometric bands is particularly challenging due to the limited spectral resolution (R$\approx$ 60) and it requires precise modelling of both the stellar continuum and the line contribution to accurately recover the total flux. Additionally, factors such as the non-uniform transmission curve of filters, potential overlap of multiple emission lines within a single filter, the intrinsic width of the emission lines and inaccuracies in the photometric redshift, can introduce significant uncertainties. To address these challenges, we adopt the following strategy for estimating the fluxes of $H{\alpha}$, $H{\beta}$ and [\ion{O}{III}].

First, we model each emission line assuming a Gaussian profile with a conservative width of $\sigma = 7$\,\AA{}, to account for thermal broadening and instrumental effects. For reference, studies of giant HII regions in the context of the $L$–$\sigma$ relation report typical $\sigma$ values of $\sim$3\,\AA{} \citep[e.g.,][]{Terlevich1981MNRAS.195..839T}. This is significantly narrower than the width assumed in our modelling, which ensures that our flux estimation is robust against potential flux losses of the wings of the emission line. After that, to guarantee that a given spectral line is fully captured by a specific filter, we  check that at least 95\% of the modelled line flux lies within the transmission curve of each filter. If this condition is met, we assume the total line flux is correctly captured.\\

To isolate the emission line flux from the underlying stellar continuum, we first estimated the continuum level using the best-fit SED model provided by \textsc{CIGALE}. This model was then convolved with the transmission curves of the J-PAS filters to obtain a realistic prediction of the continuum flux in each band. Finally, the continuum contribution was subtracted from the observed flux in the filter where the emission line was detected. Each line has certain singularities that must be considered:

\begin{itemize}
    \item H$\alpha$: The easiest line to estimate because it is largely isolated except for the nearby [\ion{N}{II}] doublet. We corrected for this contamination by assuming a typical flux ratio of [N\,\textsc{ii}]$\lambda\lambda6548,6583$/H$\alpha$ $\approx$ 0.07, which provides a reasonable approximation for EELGs. This ratio was derived from an empirical analysis of all the sample of star-forming galaxies extracted from the DESI spectroscopic dataset, as presented in Bonatto et al. (in prep.). Consistent values are obtained when using the sample of EELGs from \cite{perezMontero2021MNRAS.504.1237P}. This correction should be applied only when the signal-to-noise ratio is high enough to detect [NII]. However, we apply it to all galaxies in spite of flux may be underestimated in some cases.\\

    \item {}[\ion{O}{III}] $\lambda$5007: Ideally, we selected a filter where the line is well isolated from [\ion{O}{III}] $\lambda$4959 to avoid blending. If this is not possible, we measure the combined flux of both lines and apply the theoretical ratio [\ion{O}{III}] $\lambda$4959 = 1/2.95 × [\ion{O}{III}] $\lambda$5007 to disentangle its contribution.\\

    \item H$\beta$: It is generally straightforward, but special care must be taken in cases where the nearby [O\,\textsc{iii}]$\lambda4959$ line may contaminate the observed flux. When both H$\beta$ and [O\,\textsc{iii}]$\lambda4959$ fall within the same filter, we verify whether [O\,\textsc{iii}]$\lambda5007$ is detected in isolation in an adjacent filter. If so, we estimate its flux and assume the theoretical line ratio to correct for the contribution of [O\,\textsc{iii}]$\lambda4959$ to the blended H$\beta$+[O\,\textsc{iii}]$\lambda4959$ measurement. This allows us to recover a physically consistent estimate of the H$\beta$ flux, while also ensuring that the [O\,\textsc{iii}] lines are treated in a coherent manner. 
    
\end{itemize}

 In Fig. \ref{fig:flux}, we compare the fluxes obtained with J-PAS to those measured in the DESI/DR1 catalogue using LIME as a fitting line tool with S/N > 3. The DESI fluxes have not been aperture-corrected, as most sources are compact and the nebular emission fits within the 1.5" diameter fibre. Overall, the emission line fluxes derived from photometric bands are consistent with spectroscopic measurements within the errors ($\sigma_{NMAD}<$ 0.09 dex).

\begin{table*}[h]
    \tiny
    \centering
    \caption{Computed emission line quantities for selected galaxies.}
    \begin{tabular}{cccccccccc}
        \toprule
        \toprule
        \textbf{Tile ID} & \textbf{Number} & \textbf{Redshift} &
        \textbf{$\xi_{\mathrm{H}\alpha}$}$^{(*)}$ &
        \textbf{$\xi_{\mathrm{H}\beta}$}$^{(*)}$ &
        \textbf{F$_{\mathrm{H}\alpha}$}$^{(**)}$ &
        \textbf{F$_{\mathrm{H}\beta}$}$^{(**)}$ &
        \textbf{EW$_{\mathrm{H}\alpha}$}$^{(***)}$ &
        \textbf{EW$_{\mathrm{H}\beta}$}$^{(***)}$ &
        \textbf{EW$_{\mathrm{[O\,III]}\,5007}$}$^{(***)}$ \\
        \midrule
        8838 & 7985  & 0.17  & 24.86 $\pm$ 0.42 & 25.14 $\pm$ 0.53 & 44.89 $\pm$ 6.95 & 22.83 $\pm$ 11.79 & 318.43 $\pm$ 49.33 & 106.25 $\pm$ 54.88 & 288.67 $\pm$ 58.11 \\
        8838 & 13110 & 0.46  & --               & 24.99 $\pm$ 0.08 & --               & 83.05 $\pm$ 7.71  & --                & 148.55 $\pm$ 13.80 & 572.64 $\pm$ 20.79 \\
        8838 & 15288 & 0.398 & --               & 24.55 $\pm$ 0.57 & --               & 3.80 $\pm$ 8.82   & --                & 25.09 $\pm$ 58.32  & 468.80 $\pm$ 55.12 \\
        8838 & 18374 & 0.586 & --               & 24.80 $\pm$ 0.58 & --               & 8.91 $\pm$ 8.25   & --                & 65.80 $\pm$ 60.94  & 420.86 $\pm$ 55.43 \\
        ...  & ...   & ...   & ...              & ...              & ...              & ...               & ...                & ...                & ... \\
   
        \bottomrule
    \end{tabular}

    \vspace{2mm}
    \begin{minipage}{\textwidth}
    \small
    \textbf{Notes.}
    $^{(*)}$ in Hz erg$^{-1}$;
    $^{(**)}$ in $10^{-19}$ erg s$^{-1}$ cm$^{-2}$ \AA$^{-1}$;
    $^{(***)}$ in \AA. The entire version of this table for the full sample of EELGs is available at the CDS.
    \end{minipage}
    \label{tab:line_quantities}
\end{table*}

\subsection{Line flux equivalent widths}
\begin{figure}[h]
     \centering
     \includegraphics[width=0.9\linewidth]{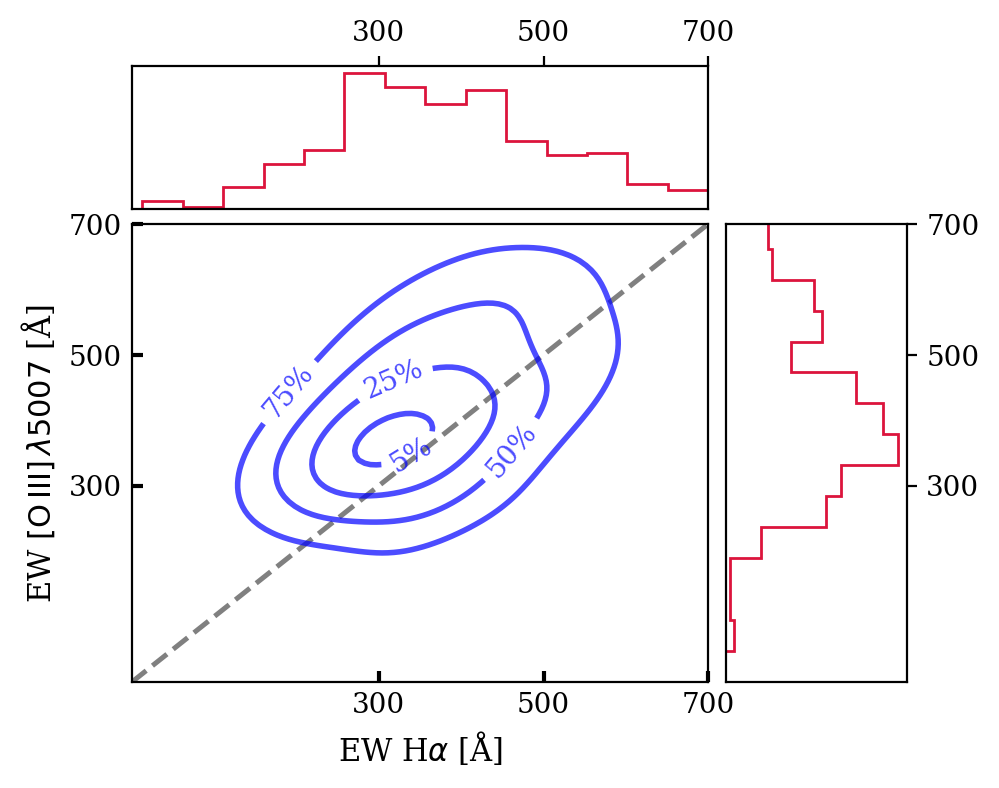}
     \caption{Contour plot of the equivalent widths of $H\alpha$ and $H\beta$, with red histograms showing the individual distributions along each axis. The contours represent the density of sources in the EW($H\alpha$)–EW($H\beta$) plane. Percentage shows the accumulative distribution and the grey line 1:1 correlation.}
     \label{fig:EW-hist}
 \end{figure}

\begin{figure}[h]
    \centering
    \includegraphics[width=0.9\linewidth]{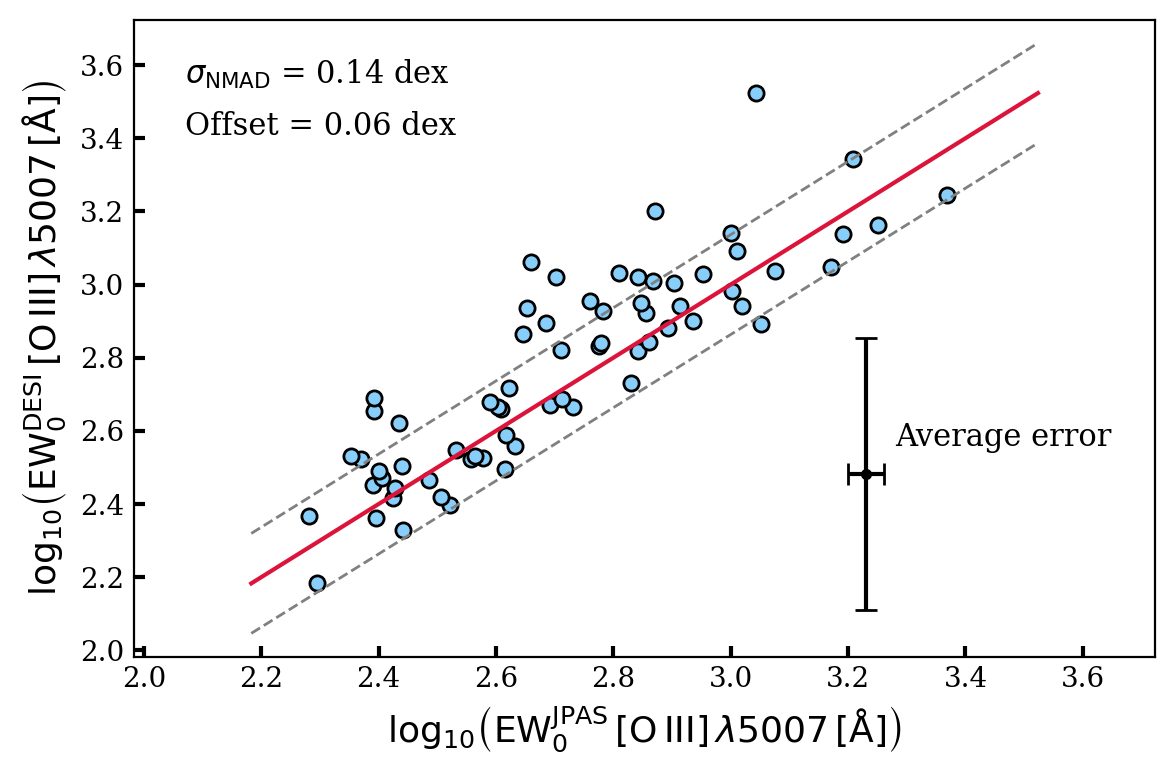}
    \caption{J-PAS EW compared with the spectroscopic EW from the DESI counterpart for [\ion{O}{III}] $\lambda$5007\, \AA\, line emission. White triangles are upper limits values. Grey lines indicates de limits of the $\pm1\sigma$ region.}
    \label{fig:ew}
\end{figure}

To compute the EW, we first convolve the synthetic stellar and nebular continuum models derived from \texttt{CIGALE} with the transmission curve of the J-PAS filters. This allows us to estimate the continuum flux within each filter. We then calculate the EW using the measured flux in the filter, following the expression:

\begin{equation}
    \text{EW} \ [\text{\AA}] = \frac{\text{Flux} - \text{Cont.}}{\text{Cont.}} \times \Delta ,
\end{equation}
 where $\Delta$ represents the filter effective width. This assumes a flat continuum across the filter and may introduce differences with respect to spectroscopic measurements. The mean EW$_o$ for the full sample is 665 \,\AA\ in  [\ion{O}{III}] $\lambda$5007  and 383 \,\AA\ in $H\alpha$. However, there are instances where the EW is above 300 \,\AA\ in [\ion{O}{III}] $\lambda$5007 but not in $H\alpha$, and viceversa. In Fig. \ref{fig:EW-hist}, the histograms for EW of [\ion{O}{III}] $\lambda$5007 and $H\alpha$ are shown. The peak observed around 300 \AA\ is a consequence of our selection criteria, which impose a minimum EW threshold as part of the methodology. This effectively cuts the original distribution. Since objects with high EWs are rare, this selection removes the low EW part of the distribution, where most objects lie, resulting in an artificial peak at 300\,\AA. Also, the threshold value in EW was imposed on non rest-frame spectra with unknown redshift at that time, so it is normal that some objects appear below that limit. In Fig.\ref{fig:ew}, we compare the pure photometric measurement of EW (S/N >3) with the values drawn for their spectroscopic counterparts. We observe a very good agreement between the two, with consistent results. The larger error bars in the spectroscopic sample are attributed to the lower continuum levels in these galaxies compared to their J-PAS counterparts. This difference in continuum strength affects the precision in the spectroscopic measurements, leading to increasing uncertainties.

\begin{figure}
    \centering
    \includegraphics[width=1\linewidth]{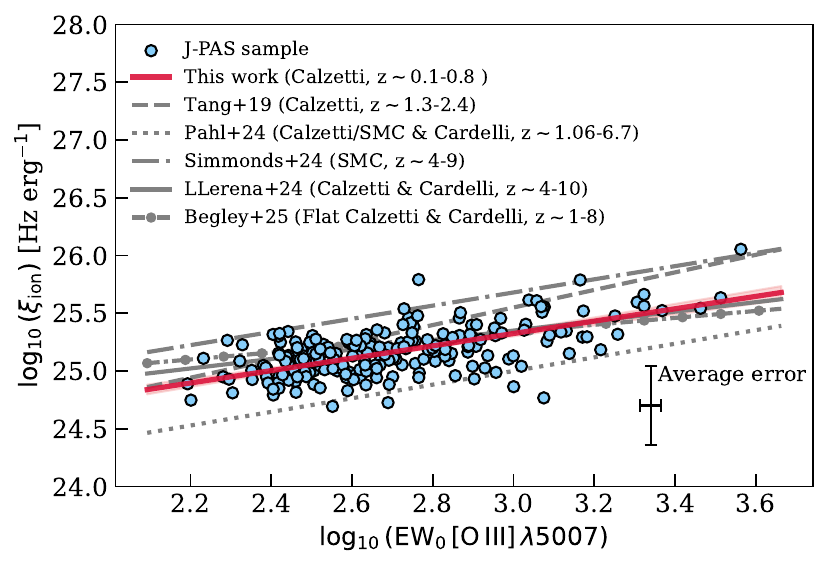}
    \caption{
Relation between the ionizing photon production efficiency, $\log(\xi_{\mathrm{ion}})$, and the rest-frame equivalent width of [O\,\textsc{III}], $\log(\mathrm{EW}_0\, \mathrm{[O\,III]})$. The blue points represent our sample of EELGs, with a typical error bar shown in the lower right region. The solid red line indicates the best-fit linear relation obtained in this work using an MCMC approach, with the shaded area showing the 1$\sigma$ confidence interval. For comparison, previous relations from the literature are also shown:\citet[dashed]{Tang2019MNRAS.489.2572T}, 
\citet[dotted]{Pahl2025ApJ...981..134P}, 
\citet[dash-dotted]{Simmonds2024MNRAS.527.6139S},  
\citet[solid]{Llerena2024arXiv241201358L}, 
\citet[solid with dot]{begley2025MNRAS.537.3245B}
}
\label{fig:xi}
\end{figure}

\section{Discussion}
\label{section5}
In this section, we investigate the ionizing photon budget of our EELG sample. 
The production of ionizing photons is dominated by very young and hot massive stars, whose presence is linked to the strength of nebular emission lines. In particular, high [O\,III]$\lambda5007$ equivalent widths trace bursts of star formation and therefore correlate with the stellar population and the ionizing photon production efficiency ($\xi_{\rm ion}$). Age is also expected to correlate with $\xi_{\rm ion}$, since younger stellar populations host larger numbers of hot, massive stars capable of producing ionizing radiation. Exploring these relationships in local EELGs provides a way of interpreting the conditions under which galaxies can contribute to the ionizing budget.

\subsection{The ionizing rate of EELGs}
$\xi_{\text{ion}}$, quantifies the number of hydrogen-ionizing photons emitted per unit of UV continuum luminosity at 1500\,\AA. It measures how efficiently young hot stars convert their ultraviolet light into ionizing radiation capable of affecting the surrounding gas. Together with the cosmic star formation rate density ($\rho_{\text{SFR}}$) and the escape fraction of Lyman continuum photons ($f_{\text{esc}}$), it is possible to estimate the rate of ionizing photons that are produced and successfully escape into the intergalactic medium \citep{roberson2015ApJ...802L..19R}. Primeval galaxies with strong bursts of star formation emit significant amounts of photoionizing radiation, suggesting that they are key agents for the reionization of the Universe \citep{Trebitsch2017MNRAS.470..224T,finkelstein2019ApJ...879...36F}. Thus, studying $\xi_{\text{ion}}$ helps us to assess the capability of EELGs, to drive cosmic reionization. In this context, EELGs in the local Universe provide valuable analogues to high-redshift systems. These galaxies often exhibit compact morphologies, low metallicities, hard radiation fields and high specific star formation rates. Moreover, several EELGs have shown direct detections of Lyman continuum leakage with significant escape fractions ($f_{\text{esc}} \gtrsim 0.1$; \citealp{Izotov2016MNRAS.461.3683I, Izotov2018MNRAS.478.4851I, Flury2022ApJ...930..126F, Flury2022ApJS..260....1F, Vanzella2016ApJ...825...41V}). We evaluate the ionizing photon production efficiency by

\begin{equation}
    \xi_{\mathrm{ion}} \ (\mathrm{Hz} \ \mathrm{erg}^{-1}) = \frac{N(H^0)}{L_{\mathrm{UV}}},
\end{equation}

where $N(H^0)$ is the ionizing photon rate in s$^{-1}$ and $L_{\rm UV}$ is the UV luminosity density at 1500\,\AA, which we estimate from \texttt{CIGALE} models assuming a filter band of 100\,\AA\ centered at 1500\,\AA.
 Just to clarify, $L_{UV}$ is computed over the entire model, so there will be a (small) contribution from nebular continuum emission included along with the stellar population. To estimate the ionizing photon rate, we use the dust-corrected H$\alpha$ luminosity from \cite{Leitherer1995ApJS...96....9L}, assuming that no ionizing photons escape the galaxy and that case B recombination applies. Additionally, we also utilize the dust-corrected H$\beta$ luminosity, as it enables us to cover a wider range of redshifts:

\begin{equation}
    \begin{aligned}
    L(H\alpha) \ (\mathrm{erg} \ \mathrm{s}^{-1}) &= 1.36 \times 10^{-12} \ N(H^0) \ (\mathrm{s}^{-1}), \\
    L(H\beta) \ (\mathrm{erg} \ \mathrm{s}^{-1}) &= 4.7 \times 10^{-13} \ N(H^0) \ (\mathrm{s}^{-1}).
    \end{aligned}
\end{equation}

This equation exhibits a mild dependence on temperature and metallicity \citep{Charlot2001MNRAS.323..887C}, and is also sensitive to the assumed IMF and stellar metallicity \citep{Atek2022MNRAS.511.4464A, Wilkins2019MNRAS.490.5359W}.
The derived value of $\xi_{\text{ion}}$ should be considered a lower limit, as we assume $f_{\text{esc}} = 0$, for instance, a radiation-bounded nebula. When the escape fraction increases, not all ionizing photons are reprocessed into Balmer emission lines, which leads to an increase in $\xi_{\text{ion}}$ by a factor of $1/(1 - f_{\text{esc}})$. We only consider $\xi_{\text{ion}}$ values reliable when Balmer lines are detected with a signal-to-noise ratio (S/N) > 3. 
\subsection{Ionizing rate vs EW (\ion{O}{III}])}
The equivalent width of [\ion{O}{III}] $\lambda$5007 has often been used as a proxy for $\xi_{\text{ion}}$. \citet{Chevallard2018MNRAS.479.3264C} established this relation using local analogues from SDSS, selected to resemble high-redshift galaxies. More recent studies have confirmed a positive correlation between EW([O~\textsc{iii}])$\lambda$5007 and $\xi_{\text{ion}}$ in high-redshift samples: \citet{Simmonds2024MNRAS.527.6139S} analysed 677 galaxies at $z \sim 4$--9 using JWST/NIRCam photometry, \citet{Llerena2024A&A...691A..59L} examined a sample of 761 galaxies at $4 \leq z \leq 10$ from various JWST surveys, and \citet{Pahl2025ApJ...981..134P} studied 163 galaxies spanning $1.06 < z < 6.7$. Additionally, \citet{Tang2019MNRAS.489.2572T} focused on a sample of extreme [O~\textsc{iii}] emitters at $z = 1.3$--2.4. In Fig.~\ref{fig:xi}, we show that this trend is consistent with our data.  We performed a linear fit to the observed correlation between $\log(\mathrm{EW}_0\, \mathrm{[O\,III]})$ and $\log(\xi_{\mathrm{ion}})$ using an MCMC approach. The resulting best-fit relation is:

\begin{align}
\log(\xi_{\mathrm{ion}})\,(\mathrm{Hz\,erg}^{-1}) &= (0.69 \pm 0.15) \times \nonumber 
\log(\mathrm{EW}_0\, \mathrm{[O\,III]})\\ & + (23.35 \pm 0.06)
\end{align}
This relation confirms a moderate positive correlation, where galaxies with stronger EW ([OIII]) tend to have higher ionizing photon production efficiencies. The slope of $\sim$0.69 is consistent with recent findings in the literature, though slightly shallower than some previous works (\citealp{begley2025MNRAS.537.3245B,Pahl2025ApJ...981..134P,Simmonds2024MNRAS.527.6139S,Llerena2024arXiv241201358L,Tang2019MNRAS.489.2572T}). This supports the idea that EW([O\,III]) can serve as a useful proxy for identifying galaxies with elevated ionizing output.\\
An important consideration in this work is the choice of extinction law. We adopted the attenuation curve from \citet{calcetti2000ApJ...533..682C}, with $R_V = 4.05$, to ensure consistency with previous studies and with the SED fitting approach implemented in CIGALE. However, several studies have suggested that a steeper attenuation law lacking the UV bump and characterized by lower values of $R_V$, typically around $R_V \sim 2.7$—may be more appropriate for compact, star-forming galaxies at high redshift \citep{Shivaei2020ApJ...899..117S, Reddy2018ApJ...853...56R, Izotov2017MNRAS.467.4118I}. Adopting an SMC-type extinction law can lead to increases of up to 0.3 dex in the derived values of $\xi_{\text{ion}}$. However, when running \texttt{CIGALE} with a steeper attenuation curve (e.g., SMC-like, with $R_V \sim 2.7$), we found that the uncertainties in the estimated $E(B{-}V)$ values increased significantly, resulting in larger uncertainties in the inferred $\xi_{\text{ion}}$. Additionally, \texttt{CIGALE} assumes a single attenuation law for both stellar and nebular emission, applying the same $E(B{-}V)$ correction to both components in the dustatt\_calzleit module. This differs from the typical spectroscopic approach, where the stellar UV continuum is corrected using Calzetti or SMC-type laws based on SED fitting, while the nebular lines are corrected using the \citet{Cardelli1989ApJ...345..245C} law together with the Balmer decrement. However, due to our instrumental constraints in particular, the inability to measure H$\alpha$ for galaxies at $z \gtrsim 0.4$ and the added uncertainties that would result from propagating errors in both H$\alpha$ and H$\beta$, we adopted the extinction derived from SED fitting to correct both the continuum and line emission. While this introduces some systematic uncertainty, it provides a consistent and homogeneous framework for estimating $\xi_{\text{ion}}$ across our full sample.
\subsection{Ionizing rate vs. age}
\begin{figure}[h]
    \centering
    \includegraphics[width=1\linewidth]{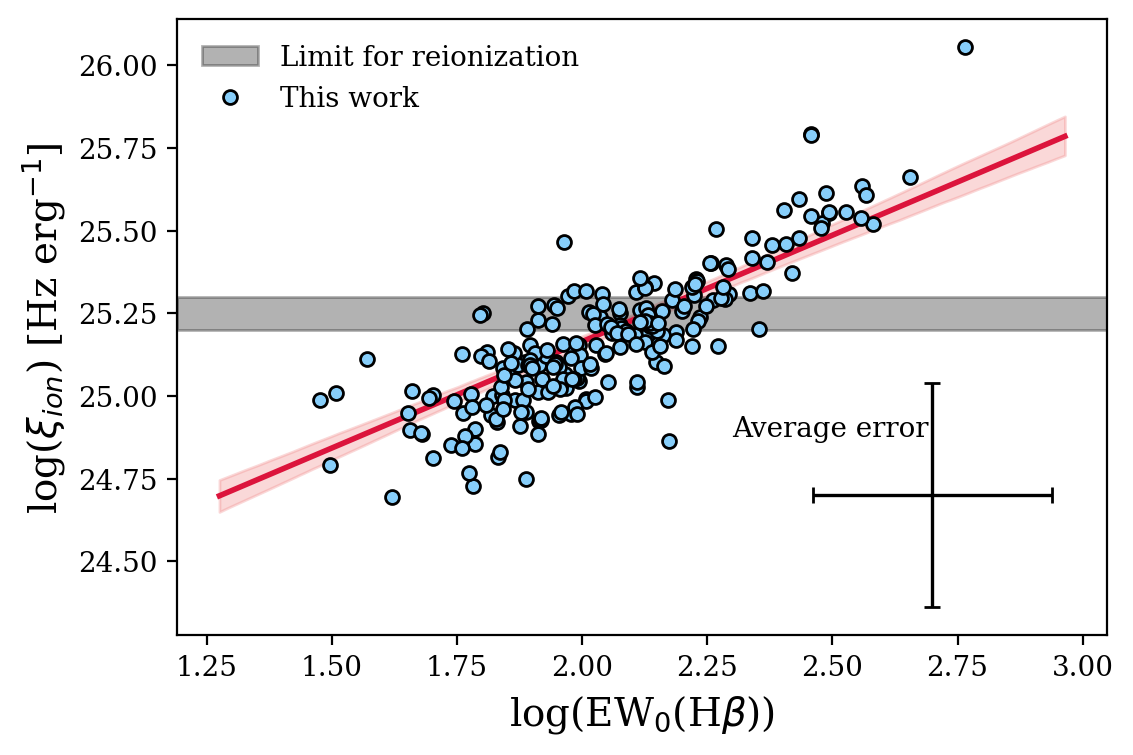}
    \caption{Relation between the ionizing photon production efficiency, $\log(\xi_{\mathrm{ion}})$, and the rest-frame equivalent width of H$\beta$, $\log[EW_0(H\beta)]$. The blue points represent our sample. The shaded grey region marks the commonly adopted threshold ($\log(\xi_{\mathrm{ion}}) > 25.3$).}

    \label{fig:hbeta}
\end{figure}

We find that $\xi_{\text{ion}}$ shows a strong correlation with both EW([OIII]) and EW(H$\beta$) (see Fig. \ref{fig:hbeta}), a trend consistent with previous studies in the literature. The correlation with EW(H$\beta$) is particularly pronounced. According to models (e.g. Starburst99, \citealp{Leitherer1999ApJS..123....3L,pystarburst2025ApJS..280....5H}), the equivalent width of H$\beta$ is a reliable tracer of the age of a starburst, reaching maximum values (greater than 200 \AA) within the first 3–4 Myr and declining rapidly afterward. In Fig.~\ref{fig:hbeta}, we illustrate the relationship between EW(H$\beta$) and the ionizing photon production efficiency. Our results strongly suggest that galaxies dominated by younger bursts exhibit the highest efficiencies. This aligns with expectations for high-redshift galaxies, where star formation tend to be stochastic and/or bursty (e.g.  \citealp{Sun2023ApJ...955L..35S,Pallottini2023A&A...677L...4P,Dressler2024ApJ...964..150D})

\begin{align}
	\log\!\left(\xi_{\mathrm{ion}}\right)\,[\mathrm{Hz\,erg}^{-1}]
	&= (0.643 \pm 0.063)\times
	\log\!\left(\mathrm{EW}_0(\mathrm{H}\beta)\right) \nonumber \\
	&\quad + (23.87 \pm 0.12)
\end{align}

The commonly adopted threshold of \(\log_{10}(\xi_{\rm ion}/\mathrm{erg^{-1}\,Hz}) \sim 25.3\) \citep{roberson2015ApJ...802L..19R} identifies galaxies capable of significantly contributing to the ionizing photon budget required for cosmic reionization. 
A fraction of the J-PAS EELGs exceed this limit. Notably, galaxies hosting the youngest bursts,are precisely those that reach or surpass this threshold, reflecting their intense production of Lyman-continuum photons. This directly links these young bursts to the cosmic reionization budget, as they can provide a significant fraction of the photons necessary to ionize the intergalactic medium.  Current evidence suggests that the \(\xi_{\rm ion}\) of these low-redshift EELGs is comparable to that of galaxies at \(z > 6\), implying little evolution of this parameter across cosmic time.

\section{Conclusions}
\label{section6}
In this work, we present a novel method for the photometric identification of EELGs based on EW, combining a classical approach, measuring the EW directly from narrow band photometry with the application of artificial intelligence techniques. The selection method relies on detecting strong emission lines, specifically those with rest frame EW([O\,\textsc{iii}]) or EW(H$\alpha$) greater than 300\,\AA, by measuring the contrast between J-PAS narrow band filters and the estimated continuum. Using data from the J-PAS Internal Data Release (IDR202406), we applied this methodology to select a sample of 917 EELGs up to a redshift of $z = 0.8$, over an area of $\approx$ 30 deg$^2$. Our selection achieves a purity of 95\%, with an estimated AGN contamination of only 5\%, and a completeness of 96\%, as determined by comparison with DESI spectroscopic counterparts. With these criteria, we find a density of 31 EELGs per square degree—nearly doubling the density reported in the miniJPAS study by \citet{iglesias2022A&A...665A..95I}, which identified 17 EELGs per square degree.\\

We carried out SED fitting using the \textsc{CIGALE} software to characterize the physical properties of the galaxies. Most of the sources have stellar masses in the range $10^7$–$10^{10}\,M_{\odot}$, with a median value of $\log(M_\star/M_{\odot}) = (8.66\pm 0.02$). Furthermore, we are able to recover a population of low-mass galaxies with stellar masses below $10^7\,M_{\odot}$, which are not accessible to DESI. This bias is probably due to the sample selection of DESI. One of the main advantages of J-PAS is the absence of selection biases, except for those inherently linked to the survey’s magnitude limit (the Malmquist bias).  We used specific templates to accurately retrieve the photometric redshifts, achieving excellent agreement with the spectroscopic redshifts that allows us to have confidence in the physical parameter derived such us line fluxes and $\xi$. Ionizing photon production is particularly important during the epoch of reionization. Nearby galaxies that exhibit physical characteristics reminiscent of early-universe systems, often referred to as local analogues, offer a unique window into the conditions that regulate ionizing photon output. Their compact structure, intense star formation, and low chemical enrichment make them the perfect systems to study what physical conditions lead to efficient ionizing photon production and how those photons might escape into the surrounding medium. Our selected J-PAS EELG sample provides a solid starting point for an unbiased study of local analogues, with most sources exceeding the minimum efficiency required to reionize the universe ( $\log \xi_\mathrm{ion}^{limit} =25.3$ ) at z > 6.\\
Our method offers an efficient way to identify EELGs using narrow-band photometry. The selected sample helps us understand the physical conditions that lead to efficient ionizing photon production in galaxies similar to those from the early universe. A spectroscopic follow-up of these sources will shed light on photon escape mechanisms and their nebular physical properties and chemical abundances. The presented catalogue will be extended as new J-PAS data become available.\\

\begin{acknowledgements}

We thank the referee for several helpful suggestions. AGA, MGO and IM acknowledge financial support from the Severo Ochoa grant CEX2021-001131-S, funded by MICIU/AEI/10.13039/501100011033. AGA also acknowledges FPI support under grant code CEX2021-001131-S-20-7. Both AGA and MGO acknowledge support from the research grant PID2022-136598NB-C32 ("Estallidos8"). MGO also acknowledges the support by the project ref. AST22\_00001\_Subp\_11 funded from the EU – NextGenerationEU. RA acknowledges support from PID2023-147386NB-I00 funded by MICIU/AEI/10.13039/501100011033 and ERDF/EU. IM acknowledges support from PID2022-140871NB-C21 funded by MICIU/AEI/10.13039/501100011033 and FEDER/UE.

RGD acknowledge financial support from the project PID2022-141755NB-I00, and the Severo Ochoa grant CEX2021-001131-S funded by MICIU/AEI/ 10.13039/501100011033.

JAFO and AE acknowledge support from the Spanish Ministry of Science and Innovation and the EU–NextGenerationEU through the RRF project ICTS-MRR-2021-03-CEFCA.
AHC and ALC acknowledge support from MCIN/AEI/10.13039/501100011033, “ERDF A way of making Europe”, and “EU NextGenerationEU/PRTR” through PID2021-124918NB-C44 and CNS2023-145339, as well as from the RRF project ICTS-MRR-2021-03-CEFCA ALC and RPT acknowledge the financial support from the European Union - NextGenerationEU through the RRF program Planes Complementarios con las CCAA de Astrof\'{\i}sica y F\'{\i}sica de Altas Energ\'{\i}as - LA4. I.B. acknowledges support from the EU Horizon 2020 programme (Marie Sklodowska-Curie Grant 101059532) and the Franziska Seidl Funding Program, University of Vienna. 

This paper has gone through internal review by the J-PAS collaboration. Based on observations made with the JST/T250 telescope and JPCam at the Observatorio Astrofísico de Javalambre (OAJ), in Teruel, owned, managed, and operated by the Centro de Estudios de Física del Cosmos de Aragón (CEFCA). We acknowledge the OAJ Data Processing and Archiving Unit (UPAD) for reducing and calibrating the OAJ data used in this work. Funding for the J-PAS Project has been provided by the Governments of Spain and Arag\'on through the Fondo de Inversiones de Teruel; the Aragonese Government through the Research Groups E96, E103, E16\_17R, E16\_20R, and E16\_23R; the Spanish Ministry of Science and Innovation (MCIN/AEI/10.13039/501100011033 y FEDER, Una manera de hacer Europa) with grants PID2021-124918NB-C41, PID2021-124918NB-C42, PID2021-124918NA-C43, and PID2021-124918NB-C44; the Spanish Ministry of Science, Innovation and Universities (MCIU/AEI/FEDER, UE) with grants PGC2018-097585-B-C21 and PGC2018-097585-B-C22; the Spanish Ministry of Economy and Competitiveness (MINECO) under AYA2015-66211-C2-1-P, AYA2015-66211-C2-2, and AYA2012-30789; and European FEDER funding (FCDD10-4E-867, FCDD13-4E-2685)"
     
\end{acknowledgements}

\bibliographystyle{aa}
\bibliography{references}

\appendix

\section{Sample completeness}

To assess the completeness of our J-PAS sample and understand potential selection effects, we explore its redshift distribution and the relation between redshift and absolute magnitude in the SDSS-$i$ band.

\begin{figure}[h]
    \centering
    \includegraphics[width=1\linewidth]{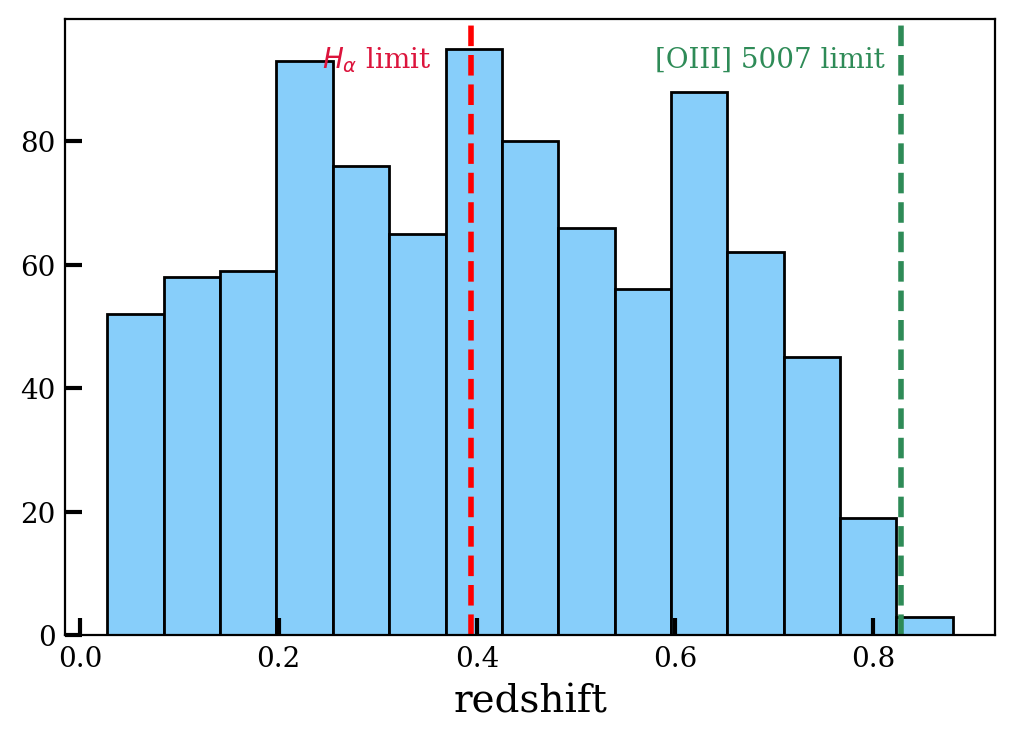}
    \caption{Redshift distribution of the selected sample. The red and green lines mark the redshift limits where the H$\alpha$ and [\ion{O}{III}] $\lambda$5007, lines, start to fall outside the observed spectral range.}
    \label{fig:z}
\end{figure}

\begin{figure}[h]
    \centering
    \includegraphics[width=0.9\linewidth]{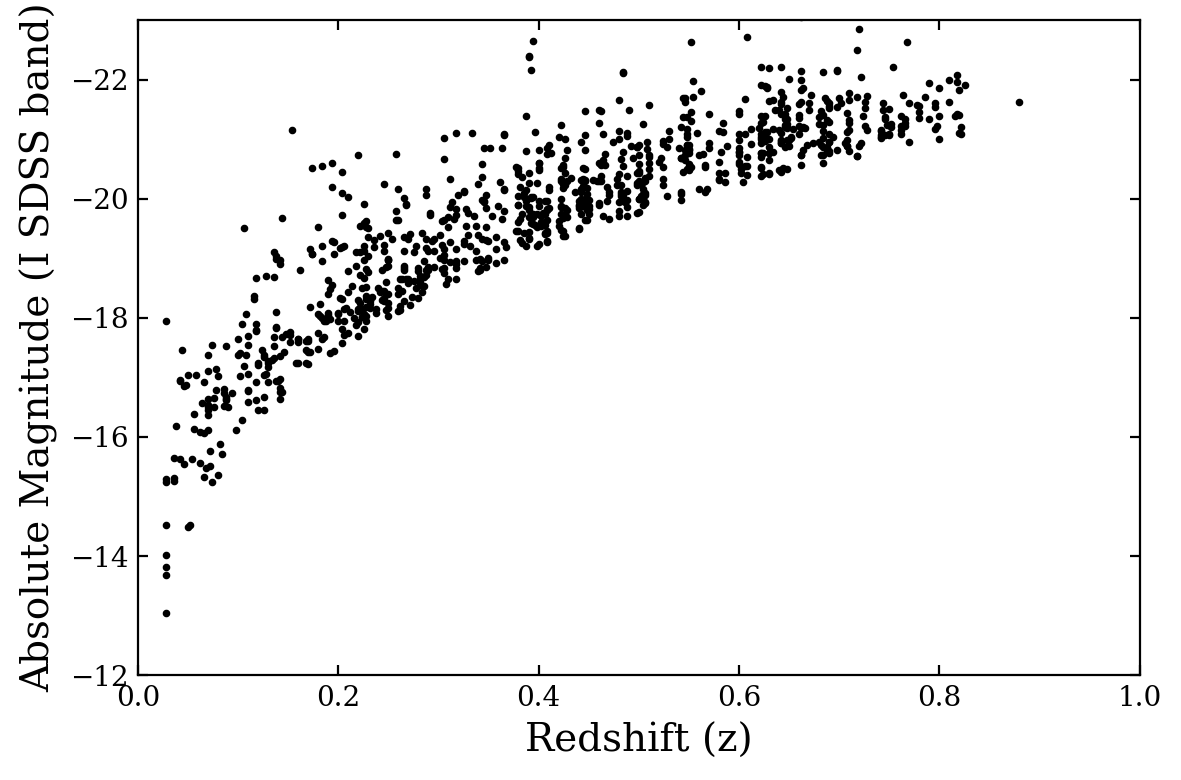}
    \caption{Distribution of galaxies in absolute magnitude in the SDSS $i$-band as a function of redshift.}
    \label{fig:mal}
\end{figure}

Figure~\ref{fig:z} shows that the redshift distribution remains approximately uniform between $z \sim 0.1$ and $z \sim 0.7$, with fluctuations consistent with Poisson noise. Key emission lines used for EELG selection, such as H$\alpha$ and [O\textsc{iii}] $\lambda5007$, shift out of the J-PAS spectral coverage at $z \sim 0.4$ and $z \sim 0.7$, respectively. However, no significant decrease in source counts is observed at these redshifts. The number of detections only shows a noticeable decline beyond $z \sim 0.7$, which is mainly driven by the increasing luminosity threshold imposed by our signal-to-noise cuts and the decreasing survey sensitivity at higher redshifts. Additionally, the increase in comoving volume with redshift helps maintain a relatively flat distribution up to this point. We also explore the distribution of absolute magnitude as a function of redshift (Fig.\ref{fig:mal}). The plot illustrate the well-known Malmquist bias: at higher redshift, the sample becomes increasingly biased towards intrinsically brighter galaxies, since fainter sources fall below the survey’s detection limit. This effect is typical in flux-limited samples and must be considered when analysing the physical properties and number densities of EELGs across redshift.

\section{Neural network architecture Details}
\label{ApendixNN}

This appendix provides a detailed description of the neural network architecture used in the models. The structure consists of three parallel branches, each designed to process a distinct type of input: 1D photometric spectra, full-frame 2D galaxy images, and a localized image region centered on the galaxy’s emission line, intended to identify potential cosmetic artifacts. The individual layers used in this model are summarized in Table~\ref{tab:nn_layers}.

\subsection*{Photometric branch}

This branch handles 1D vectors representing the galaxy’s photometric fluxes across multiple bands. It begins with a 1D convolutional layer, \textit{Conv1D(64, 3)}, which scans along the spectral sequence to detect local patterns in the flux distribution. A \textit{MaxPooling1D(2)} layer follows, which downsamples the signal to reduce dimensionality while preserving important features. A second convolutional layer, \textit{Conv1D(64, 3)}, captures more abstract spectral patterns. The output is then flattened and passed through a sequence of dense (fully connected) layers: \textit{Dense(128)}, followed by \textit{Dropout(0.3)} for regularization, and two smaller layers with 64 and 32 units. These layers help the model combine and transform spectral information into a compact feature representation.

\subsection{Image branch}

The image branch processes a 2D cutout of the galaxy. It starts with a convolutional layer, \textit{Conv2D(64, 3×3)}, which extracts spatial features from the image. A \textit{MaxPooling2D(2×2)} layer then reduces the spatial resolution. A second convolutional layer, \textit{Conv2D(32, 3×3)}, refines these features, followed by another \textit{MaxPooling2D}. The resulting feature maps are flattened and passed through \textit{Dense(128)} and \textit{Dropout(0.3)}, then through \textit{Dense(64)}, another \textit{Dropout(0.3)}, and finally \textit{Dense(32)}. This branch is primarily responsible for capturing the galaxy’s morphology — such as its compactness, elongation, symmetry, and surface brightness profile.

\subsection{Cosmetic branch}

The cosmetic branch receives a small image patch centered on the region where the galaxy’s emission line is located. This focused view allows the model to identify local anomalies — such as cosmic rays, hot pixels, or detector defects — that may affect the spectral measurement. The structure consists of a \textit{Conv2D(64, 3×3)} layer, followed by \textit{MaxPooling2D(2×2)}, a \textit{Flatten} operation, and two fully connected layers: \textit{Dense(64)}, \textit{Dropout(0.3)}, and \textit{Dense(32)}.

\subsection{Fusion and output layers}

After the feature extraction in each branch, the outputs are concatenated into a single combined vector. This joint representation is passed through a stack of dense layers: \textit{Dense(128)}, \textit{Dropout(0.3)}, \textit{Dense(64)}, \textit{Dense(32)}, and \textit{Dense(16)}. These layers allow the model to integrate the information from different inputs and produce the final prediction through the output node.

\begin{figure*}
    \centering
    \includegraphics[width=1\linewidth]{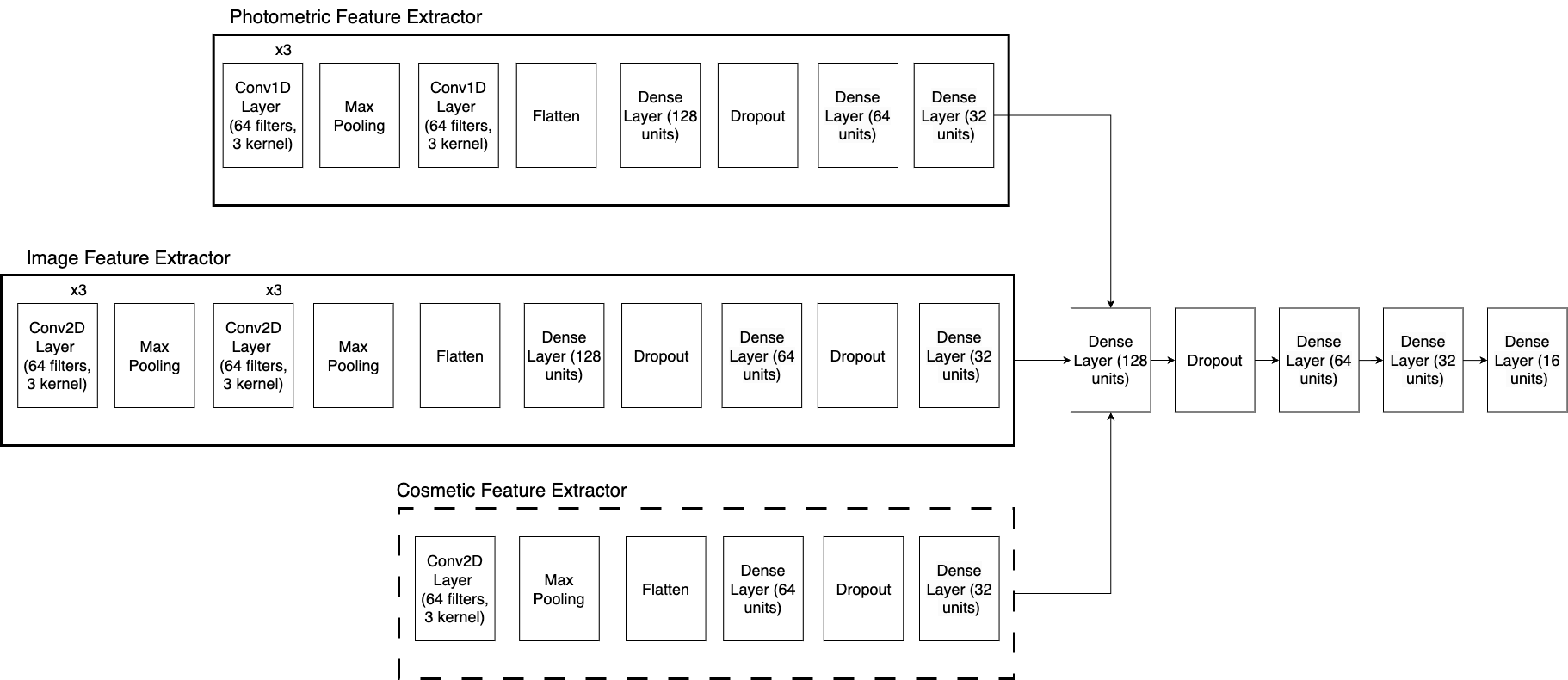}
    \caption{Model architecture consisting of three parallel branches designed for multimodal data integration: (1) a spectral branch (upper), which processes the photometric flux through 1D convolutional layers; (2) an image branch (middle), which analyses the galaxy image using 2D convolutions to extract morphological features; and (3) a defect-detection branch (lower), which processes localized image regions around the emission-line position to identify cosmetic artifacts. Each branch applies a combination of convolutional, pooling, and dense layers, with dropout regularization to prevent overfitting. The outputs from the active branches are concatenated and passed through a shared set of fully connected layers, followed by a final sigmoid activation that produces the output label. For the primary EELG classifier ($P_0$), only the spectral and image branches (upper and middle) are used. For the cosmetic-defect classifier ($P_1$), all three branches are active, including the additional defect-detection branch. A detailed description can be found in appendix \ref{ApendixNN}.
}    
    
    \label{fig:model}
\end{figure*}

\begin{table*}
    \tiny
    \centering
    \caption{Description of the neural network layers and their role in the model architecture.}
    \begin{tabular}{llp{11cm}}
        \toprule
        \toprule
        Layer Type & Acronym & Description \\
        \midrule
        1D Convolution & Conv1D($n$, $k$) & Applies $n$ filters of size $k$ across 1D input sequences (e.g., photometric spectra), enabling local pattern extraction such as flux variations between adjacent bands. \\
        2D Convolution & Conv2D($n$, $k \times k$) & Applies $n$ 2D filters over an image to extract spatial features. In this context, it helps capture galaxy morphology including compactness, shape, elongation, and brightness gradients. \\
        Max Pooling (1D) & MaxPooling1D($p$) & Reduces the dimensionality of 1D data by selecting the maximum value over a pooling window of size $p$. Helps retain prominent spectral features while reducing computational cost. \\
        Max Pooling (2D) & MaxPooling2D($p \times p$) & Downsamples 2D feature maps by selecting the maximum value within each $p \times p$ window. Reduces spatial resolution while preserving key visual features. \\
        Flatten & — & Converts multi-dimensional data (e.g., tensors) into a 1D vector suitable for input to fully connected layers. \\
        Fully Connected & Dense($n$) & A dense layer with $n$ neurons that combines and transforms input features using learned weights. Used for nonlinear integration of extracted features. \\
        Dropout & Dropout($p$) & Randomly disables a fraction $p$ of neurons during training to prevent overfitting and encourage generalization. \\
        Concatenation & — & Merges multiple input vectors into a single unified feature representation, used here to combine the outputs of the three branches. \\
        \bottomrule
    \end{tabular}

    \vspace{2mm}
    \label{tab:nn_layers}
\end{table*}
\subsection{Overtraining}

Given the relatively small size of the training dataset and the complexity of the neural network, overfitting was a major concern. To mitigate this, we employed several strategies. First, the dataset was split into distinct training and test subsets, ensuring the model was evaluated on unseen data. Second, dropout layers with a rate of 30\% were applied extensively throughout the network, effectively reducing co-adaptation between neurons and improving generalization. Third, early stopping was implemented based on the validation loss, stopping training once no further improvement was observed.

The learning curves for both training and test loss exhibit a consistent and monotonic decrease without the divergence typically associated with overfitting. This behaviour suggests that the chosen regularization strategies, particularly the dropout layers, are successfully preventing the model from memorizing the training data while maintaining strong predictive performance.

\begin{figure}[h]
    \centering
    \includegraphics[width=1\linewidth]{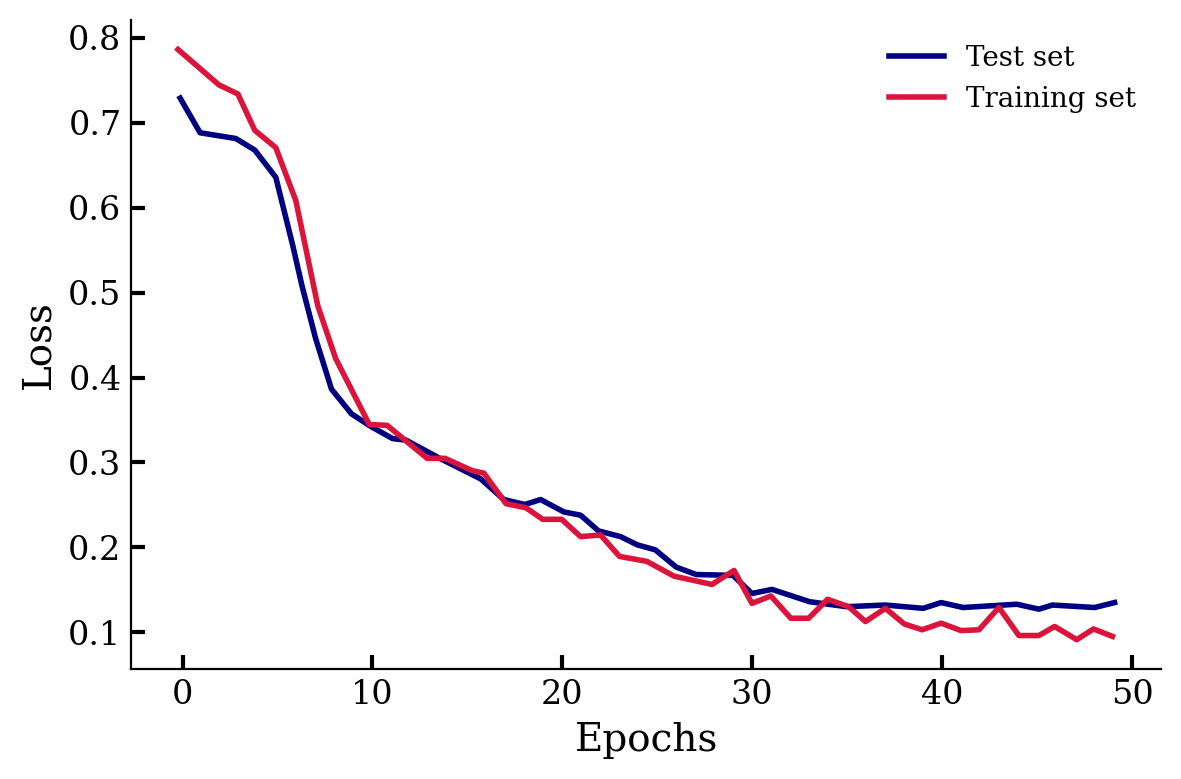}
    \caption{Test loss (blue) and training loss (red) as a function of training epochs. Both decrease monotonically without signs of overfitting.}
    \label{fig:placeholder}
\end{figure}

\begin{figure}[h]
    \centering
    \includegraphics[width=0.9\linewidth]{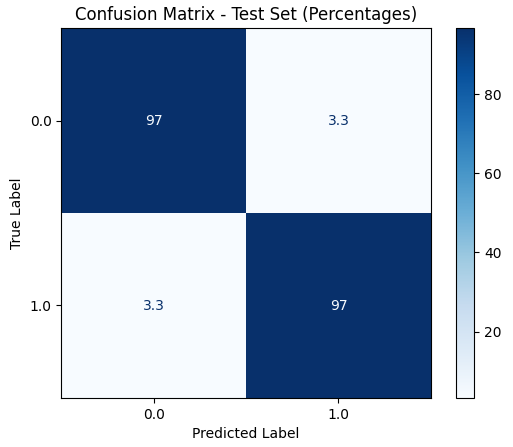}
    \caption{Confusion matrix illustrating balanced prediction rates for true positives, true negatives, false positives, and false negatives, each occurring with similar percentages, indicating a well-calibrated classifier without bias toward any class.}

    \label{fig:placeholder}
\end{figure}

\subsection{Confusion matrix analysis}

The confusion matrix provides a detailed breakdown of the classification performance by showing the counts or percentages of true positives (TP), true negatives (TN), false positives (FP), and false negatives (FN). In this particular case, the matrix reveals that the model predicts each of these outcomes with  the same percentage. This balance indicates that the classifier does not favor any particular class and treats positive and negative predictions with similar accuracy. Specifically, the true positives and true negatives show that the model correctly identifies positive and negative cases at similar rates. At the same time, the false positives and false negatives, which represent the model’s errors in classification, also occur with comparable frequency. This symmetric distribution suggests that the model's decision boundary is well-calibrated and does not introduce significant bias toward one class.
\newpage

\section{Cigale parameters}
\begin{table}[h]
    \tiny
    \centering
        \caption{\texttt{CIGALE} Parameters}
        \begin{tabular}{cc}
        \toprule
        \toprule
            \textbf{Stellar Parameters} &  \\ \midrule
            $\tau_{\mathrm{main}}$ {[}Myr{]} & 5, 10, 20, 50 \\ \rule{0pt}{2.3ex}
            $\tau_{\mathrm{burst}}$ {[}Myr{]}      &  0.5, 0.8, 1, 3 \\ \rule{0pt}{2.3ex}
            Age main {[}Myr{]} & 30, 50, 100, 200, 500        \\ \rule{0pt}{2.3ex}
            burst\_age {[}Myr{]} & 1, 3, 6, 10, 12 \\ \rule{0pt}{2.3ex}
            f\_burst& 0.1, 0.15, 0.25, 0.3, 0.5 \\
            \midrule
            \textbf{Charlot \& Bruzual (2019)} &  \\ \midrule 
            IMF & Chabrier \\ \rule{0pt}{2.3ex}
            metallicity & 0.0001, 0.001, 0.004, 0.008\\ \rule{0pt}{2.3ex}
            Upper IMF limit [M$_\odot$] & 100\\

            \midrule 
            \textbf{Nebular parameters} &  \\ \midrule 
            z\_gas & 0.0001, 0.001, 0.004, 0.008   \\ \rule{0pt}{2.3ex}
            $\log{U}$  & -3.5, -3.0, -2.5, -2.0, -1.5   \\ \rule{0pt}{2.3ex}
            f$_{esc}$, f$_{dust}$ & 0\\

            \midrule
            \textbf{Extinction parameters} &  \\ \midrule
            $E(B-V)_\mathrm{young}$  & 0.1, 0.2, 0.3, 0.4  \\ \rule{0pt}{2.3ex}
            $E(B-V)_\mathrm{old\_factor}$    & 0.44, 1     \\ 
            \noalign{\smallskip}

             \midrule
            \textbf{Dust parameters} &  \\ \midrule
            AGN Fraction  & 0  \\ \rule{0pt}{2.3ex}
            $\alpha$    & 0.5    \\ 
            \noalign{\smallskip}
            
            \bottomrule
        \end{tabular}
        
    \label{tab:sed_parameters}
    \end{table}

\end{document}